\documentclass[useAMS,usenatbib,usegraphicx]{mn2e}
\usepackage{times,amsmath}

\DeclareMathAlphabet{\mathsfbf}{OT1}{phv}{b}{n}

\newcommand\mat[1]{\ensuremath{\mathsfbf{#1}}}
\newcommand\vect[1]{\ensuremath{\mathbf{#1}}}
\newcommand\uvect[1]{\ensuremath{\mathbf{\hat{#1}}}}

\begin{document}
\title[Detection of satellite remnants with Gaia]
{Detection of satellite remnants in the Galactic Halo with \textit{Gaia} -- I.
The effect of the Galactic background, observational errors and sampling}

\author[A.G.A. Brown, H.M. Vel\'azquez \& L.A. Aguilar]%
{Anthony G.A. Brown$^1$\thanks{e-mail: brown@strw.leidenuniv.nl}, Hector
  M. Vel\'azquez$^2$ and Luis A. Aguilar$^2$\\
$^1$Sterrewacht Leiden, P.O. Box 9513, 2300 RA Leiden, The Netherlands\\
$^2$Instituto de Astronom{\'\i}a, UNAM, Apartado Postal 877, Ensenada, 22800,
Baja California, Mexico}

\maketitle

\begin{abstract}
We address the problem of identifying remnants of satellite galaxies in the
halo of our galaxy with \textit{Gaia} data. The \textit{Gaia} astrometric
mission offers a unique opportunity to search for and study these remnants
using full phase-space information for our Galaxy's halo. However, the
remnants have to be extracted from a very large data set (of order $10^9$
stars) in the presence of observational errors and against a background
population of Galactic stars. We address this issue through numerical
simulations with a view towards timely preparations for the scientific
exploitation of the \textit{Gaia} data.

We present a Monte Carlo simulation of the \textit{Gaia} catalogue with a
realistic number of entries. We use a model of the galaxy that includes
separate light distributions and kinematics for the bulge, disc and stellar
halo components.  For practical reasons we exclude the region within Galactic
coordinates: $-90^\circ\le\ell\le90^\circ$ and $-5^\circ\le b\le5^\circ$,
close to the Galactic plane and centre. Nevertheless, our catalogue contains
$3.5\times 10^8$ stars. No interstellar absorption has been modelled, as we
limit our study to high Galactic latitudes.

We perform tree code $10^6$--body simulations of satellite dwarf galaxies in
orbit around a rigid mass model of the Galaxy. We follow the simulations for
$10^{10}$ years. The resulting shrinking satellite cores and tidal tails are
then added to the Monte Carlo simulation of the \textit{Gaia} catalogue. To
assign photometric properties to the particles we use a Hess diagram for the
Solar neighbourhood for Galactic particles, while for the dwarf galaxy
particles we use isochrones from the Padova group. When combining the Milky
Way and dwarf galaxy models we include the complication that the luminosity
function of the satellite is probed at various depths as a function of
position along the tidal tails. The combined Galaxy and satellites model is
converted to a synthetic \textit{Gaia} catalogue using a detailed model for
the expected astrometric and radial velocity errors, depending on magnitude,
colour and sky position of the stars.

We explore the feasibility of detecting tidal streams in the halo using the
energy versus angular momentum plane. We find that a straightforward search in
this plane will be very challenging. The combination of the background
population and the observational errors will make it difficult to detect tidal
streams as discrete structures in the $E$--$L_z$ plane. In addition the
propagation of observational errors leads to apparent caustic structures in
the integrals of motion space that may be mistaken for physical entities. Any
practical search strategy will have to use a combination of pre-selection of
high-quality data and complementary searches using the photometric data that
will be provided by \textit{Gaia}.
\end{abstract}

\begin{keywords}
Galaxy: formation, structure -- galaxies: interactions -- methods: numerical
\end{keywords}

%%%%%%%%%%%%%%%%%%%%%%%%%%%%%%%%%%%%%%%%%%%%%%%%%%%%%%%%%%%%%%%%%%%%%%%%%%%%

\begin{table*}
\begin{minipage}{126mm}
\caption{Scientific capabilities of the \textit{Gaia} mission. Listed are the
  survey parameters and the measurements with their expected accuracies. Note
  that $\mu$arcsec stands for micro-arcsecond. The numbers in this table are
  from the \textit{Gaia} Concept and Technology Study Report
  \citep[see][]{ESA2000,Perryman2002}}
\label{tab:sciencecap}
\begin{tabular}{ll|ll}
\hline
\multicolumn{2}{c}{Survey parameters} & \multicolumn{2}{c}{Measurements and accuracies}\\
\hline
Magnitude limit   & 20--21 mag             & Astrometry & 4 $\mu$arcsec at $V=10$ \\
Completeness      & 20 mag                 &            & 11 $\mu$arcsec at $V=15$ \\
Number of objects & 26 million to $V=15$   &            & 160 $\mu$arcsec at $V=20$ \\
                  & 250 million to $V=18$  & Photometry & 4 broad band to $V=20$ \\
                  & 1000 million to $V=20$ &            & 11 medium band to $V=20$ \\
Observing programme & On-board and unbiased & Radial velocities & 1--10
km~s$^{-1}$ at $V=16$--$17$\\
\hline
\end{tabular}
\end{minipage}
\end{table*}

\section{Introduction}

Current cosmological models for the formation of the large scale structure of
the Universe envisage the formation of large luminous galaxies as a long
process of merging of smaller structures \citep{KauffmannWhite1993}. Although
the details that lead from the merging of dark haloes to the formation of
luminous galaxies are quite complicated and currently addressed only through
the use of semi--analytical models
\citep{Kauffmann1993,Somerville1999,Cole2000}, it is clear that such a
formation scenario must have left an imprint on the halo of luminous galaxies
like ours, where the long dynamical time scales preserve for a long time the
remnants of old mergers \citep{LyndenBell1976}. Although with low statistical
weight due to the small number of stars involved, for a long time there has
been increasing evidence of the existence of substructure in the stellar halo
of our galaxy
\citep{Eggen65,Rodgers84,Ratnatunga85,Dionidis89,Arnold92,Majewski96}. This
trend culminated with the discovery of the Sagittarius Dwarf galaxy by
\citet{Ibata94} and the discovery of the Canis Major dwarf
\citep{Newberg2002,Martin04}. These satellite galaxies are in the process of
being torn apart by the tidal field of our Galaxy.

The identification and study of substructure in our stellar halo is of
paramount importance, not only because it traces the formation of our Galaxy,
but because it allows us to probe the clustering of dark matter at small
spatial scales \citep[for a recent review see][]{Freeman2002}. This clustering
is a crucial clue to the nature of its constituents: The nature of the dark
matter sets the scale at which dark matter can clump
\citep{Bond83,Navarro97,Klypin99,Moore99}. A popular current candidate, the
axion, is ``cold'' enough that it can clump at scales of dwarf galaxies.
Although evidence for dark matter within dwarf galaxies orbiting our galaxy
has been put forward \citep{Irwin95}, this evidence is weakened by the
unfortunate degeneracy between potential well depth and orbital eccentricity
in the dynamical models \citep{Wilkinson02}. Identification and modelling of
several satellite systems in the process of being torn apart by the tidal
field of our galaxy can help to disentangle this degeneracy, as the rate of
erosion of the satellite depends more importantly on the depth of its own
potential well than on the details of its internal dynamics \citep[Roche Lobe
argument,][]{MB81}.

Several authors have pointed out that the thick disc of our galaxy may have
been produced by a merger event around 10 Gyr ago
\citep*{Gilmore85,CLL89}. \citet{Wyse99} has pointed out that this corresponds
to a redshift of $z\sim 2$. By probing the structure that exists within the
halo of our galaxy, we are doing `in situ' cosmological studies of events that
took place long ago. Recent observations \citep*{Stockton04,Tecza04} have shown
evidence for the existence of luminous galaxies at large redshifts ($z\sim
2.5$) that seem to be old, massive and have large metallicity. Their existence
in large numbers, if confirmed, poses a challenge to the standard cosmological
scenario of bottom-up formation. It is clear that it is of paramount
importance to study the stratigraphy of the halo of our galaxy to try to
discern what the assemblage process was by which our galaxy arose.

Great progress has recently been made in the study of sub-structure in the
halo of the Milky Way through dedicated surveys of halo giants and other
tracer populations \citep[e.g.,][]{Morrison2000,Majewski2000,Palma2003}. In
addition large scale photometric surveys, such as SDSS and 2MASS
\citep{York2000,Cutri2003} which cover the entire sky or large parts thereof,
have been exploited to explore the halo substructure
\citep[e.g.,][]{Newberg2002,Martin04,RochaPinto2004}. However all these
surveys provide at most one component of the space motion for a very small
fraction of the stars surveyed and often they only probe a small fraction of
the halo. To definitively unravel the structure and formation history of the
Galactic halo (and the Galaxy as a whole) detailed phase-space information is
required of a significant fraction of the entire volume of the Galactic halo
\citep{Freeman2002}. This requires a very high precision astrometric survey
done over the whole sky, complemented with radial velocities and photometry.

In 2000 October the \textit{Gaia} satellite mission was selected as the European Space
Agency's sixth Cornerstone mission to be launched no later than 2012. The main
scientific aim of the \textit{Gaia} mission is to map the structure of the Galaxy and
unravel its formation history by providing a stereoscopic census of 1 billion
stars throughout the Galaxy. This will be achieved by measuring accurate
positions, parallaxes and proper motions for all stars to a limiting magnitude
of $V=20$. To complement the astrometric information radial velocities will be
collected for all stars brighter than $V\simeq17$. In addition photometry will
be measured for all stars; four broad bands will be employed and about 11
intermediate bands, which will provide detailed astrophysical information for
each detected object. \textit{Gaia} will employ an on-board detection scheme allowing
it to measure every detected object brighter than $V=20$. This ensures an
unbiased and complete survey of the Galaxy. The scientific capabilities of
\textit{Gaia} are summarized in Table~\ref{tab:sciencecap}. For details see
\citet{Perryman2002}, \citet{Perryman2001}, and \citet{ESA2000}.

These measurement accuracies can be put into perspective by using Galaxy
models to calculate the corresponding distance and tangential velocity
accuracies \citep[see][]{Perryman2002}: 21 million stars will have distances
with a precision better than 1 per cent, 116 million better than 5 per cent
and 220 million better than 10 per cent; 84 million stars will have tangential
velocities with a precision better than $1$~km~s$^{-1}$, 210 million better than
$3$~km~s$^{-1}$, 300 million better than $5$~km~s$^{-1}$, and 440 million better than
$10$~km~s$^{-1}$.

In view of this kind of stereoscopic data becoming available for the Galaxy in
the future, several authors have made simulations of the process of accretion
and tidal disruption of satellite galaxies in the halo of our galaxy, with the
goal of identifying the characteristics of the resulting debris
\citep*[e.g.][]{Oh1995, Johnston1996,Meza2005}. \citet{Helmi1999}, and
\citet{Helmi2000}, in particular, have studied the problem of identifying the
resulting substructure in the stellar halo with information from a new
generation of astrometric space missions. They conclude that a promising
technique is to search for this substructure using conserved dynamical
quantities: energy and angular momentum. They performed simulations of the
destruction and smearing of satellite galaxies in the halo of our galaxy and
plotted the remnants in the $(E,L,L_z)$ space. They applied a simple model of
the astrometric errors expected in the FAME, DIVA and \textit{Gaia} missions, that
depends on the apparent magnitude alone. The errors for the \textit{Gaia} mission will
depend also on the Ecliptic coordinates and the colour of stars, and this can
introduce additional correlations between the astrometric variables. In
addition \citet{Helmi2000} only considered relatively bright and nearby
($V\la15$, $d\la6$~kpc) stars when constructing their $E$--$L$--$L_z$
diagrams. However, when searching for substructure throughout the whole volume
occupied by the Galaxy's halo, the majority of stars will be much fainter (and
much further away) and have correspondingly worse astrometric errors (see
Table~\ref{tab:sciencecap}) which will result in much more smearing in the
$(E,L,L_z)$ space. Finally, previous investigations have not taken into
account the problem of contamination by the Galactic background. It is
expected that, in any given field, the signature of an accretion event will
not dominate the corresponding star counts and a method should be devised to
identify the proverbial `needle in the haystack'.

The goals of this work are to:
\begin{enumerate}
\item[(1)] understand, through simulations, how one can detect and trace the
debris of disrupted satellites in the halo of our galaxy using direct
measurements of phase-space, as will be provided by \textit{Gaia}.
\item[(2)] include a realistic model of the Galactic background population
against which the remnants have to be detected. This includes a more elaborate
model of the observational errors expected for \textit{Gaia}.
\item[(3)] gain practical experience with analysing and visualizing the
enormous volume of multi-dimensional information that will be present in the
\textit{Gaia} data base.
\end{enumerate}

For the Galaxy model we have been careful to use Monte Carlo simulations that
result in the expected number of stars in the \textit{Gaia} catalogue to avoid
problems owing to under sampling. Our Galactic model includes spatial,
photometric and kinematic information, as all these factors affect the final
astrometric precision. For the simulated satellites, we tackle the problem of
probing the satellite luminosity function at varying depths as a function of
distance to the observer, as we move along the tidal streamer.  We present a
method to avoid loosing too many $N$-body particles from the resulting
simulated survey, while keeping a realistic model of the depth of the
observations.  At the present time we have not incorporated the effects of
dynamical friction, and this remains the main factor not yet taken into
account. Simulations like those presented by \citet{Meza2005} suggest that
this may be an important effect that further smears the expected signal in the
$E$ versus $L_z$ plane. We plan to incorporate this effect in the near future,
however, our incorporation of the Galactic background, the more realistic
astrometry errors and the addressing of the sampling details, introduces
enough new elements that publishing these results is warranted.

In Section~\ref{sec:galmod} we present our photometric and kinematic model for
the luminous components of the Galaxy and discuss the problems that have been
tackled to ensure a realistic modelling of the magnitude limited probing of
\textit{Gaia}. Section~\ref{sec:nbody} presents the details of the $N$-body
simulations that we performed to study the process of tidal dissolution of the
satellites orbiting in the potential of the Milky Way
galaxy. Section~\ref{sec:modelgaia} presents the simulated \textit{Gaia}
survey in detail, describing the model used for the observational errors and
the way the $N$-body information has been incorporated into the simulated
survey. In Section~\ref{sec:results} we discuss the appearance of the
satellites in the $E$ versus $L_z$ plane, emphasizing the effects of including
a background population and a more detailed error model. Finally,
Section~\ref{sec:future} presents our conclusions and directions for future
work.

%%%%%%%%%%%%%%%%%%%%%%%%%%%%%%%%%%%%%%%%%%%%%%%%%%%%%%%%%%%%%%%%%%%%%%%%%%%%%%%

\section{Monte Carlo Model of the Galaxy}
\label{sec:galmod}

The Galaxy simulations aim at providing a synthetic \textit{Gaia} catalogue
consisting of all stars on the sky brighter than $V=20$. This catalogue will
consist of a very large number of stars ($\sim10^9$) comprising a mix of
various populations. The main Galactic components will be stars in the bulge,
disc and halo. Superposed will be the satellites presently orbiting the Galaxy
(such as the Magellanic Clouds) as well as debris streams of satellites that
have been accreted and have not yet completely mixed with the population of
Galactic stars that was in place before the accretion events.

We have simulated in detail the disruption of satellites on various orbits
around the Galactic centre, assuming a smooth Galactic potential consisting of
a disc, bulge and halo component (see Section~\ref{sec:nbody}). This smooth
component corresponds to the above mentioned population of Galactic stars that
was in place before the accretion event started. It is against this (vastly
larger) `background' population of stars in phase-space that one has find the
tidal debris. One of our main concerns here is a realistic simulation of this
background in the \textit{Gaia} catalogue.

We can approach this problem in two ways. (a) We can assume that the
populations of stars constituting the bulk of our galaxy is already thoroughly
mixed, both spatially and kinematically, so that they form spatially smooth
components describable with simple kinematics (rotation plus velocity
dispersions). Here we implicitly assume that the bulk of our galaxy was formed
very early on in some form of monolithic collapse or that the traces of any
mergers resulting in the formation of the Galaxy have been wiped out in the
phase mixing process. (b) The hierarchical paradigm for galaxy formation
assumes that the first structures to form in the Universe are small galaxies
which through many mergers give rise to the large galaxies such as our
own. Depending on the merger history of the Galaxy it could thus well be that
the halo is not smooth but consists of a superposition of many merged small
galaxies which will lead to a lumpy phase-space structure. In this case the
bulk of the Galaxy would consist of smooth bulge and disc populations but a
halo in which the stars are clumped both spatially and kinematically.

These two options for modelling the background populations will translate in
to different distributions of the stars in phase-space. We want to stress here
that our main concern is not to build an accurate model of the Galaxy but to
use the model we have to obtain a distribution of stars in phase-space that
captures the essential features of any population that has been in place for
some time before the accretion events that we wish to find. These features are
(1) that these stars occupy the entire volume of the Galaxy according to a
density distribution that is much smoother than that of the debris of recent
accretion events; and (2) that these stars have mixed to a degree where their
kinematics is much simpler (rotation plus dispersion) than that of the
accretion debris.

Thus, although model (b) seems to be supported by recent observational
evidence we chose model (a) for simplicity. Certainly in the outer reaches of
the Galactic halo this may lead to unrealistically smooth distributions of
stars in phase-space. However, an analysis by \citet[][based on SDSS
data]{Newberg2002} of the spatial distribution of stars near the turn-off in
the Hertzsprung-Russell diagram clearly shows the presence of a smooth
component of the halo. Moreover, in the inner parts of the Galaxy (near the
Sun's position) the approximation of our Galaxy as consisting of a smooth set
of stellar populations will be valid even if the Galactic halo consists of
superpositions of streams of many accreted satellites. This was argued
recently by \citet*{Helmi2002} on the basis of the analysis of high-resolution
cosmological simulations of the formation of galactic haloes. They showed that
in these simulations the phase-space distribution of dark matter particles in
the Solar vicinity is very smooth spatially, with the velocity ellipsoids
deviating only slightly from Gaussian.

Finally we note that one could use these same cosmological simulations to
build a more realistic Galaxy model, although one is then faced with the
problem of translating the phase-space parameters of $\sim10^7$ dark matter
particles into those of $\sim10^9$ stars.

We use two different Galaxy models in this paper. The Monte Carlo model
generates a distribution of stars (i.e., luminous objects) for the three
components of the Galaxy, and these stars make up the bulk of the entries
in our simulated \textit{Gaia} catalogue. We will refer to this model as the `light'
model of the Galaxy. For the $N$-body simulations of the satellites we need a
Galactic model that includes both the luminous and dark matter. This model is
represented by a gravitational potential only (see Section~\ref{sec:nbody})
and we will refer to it as the `mass' model. We now discuss the Galactic light
model and how we implemented its Monte Carlo realization in practice.

\begin{table}
\caption{The three spatial components of the Monte Carlo model of the
  luminosity distribution of the Galaxy. Their analytical forms (except for
  the normalizations) are given as well as the length scales involved (in
  kpc). The variables $r$ and $(R,z)$ indicate spherical and cylindrical
  coordinates respectively, centred on the Galactic centre.}
\label{tab:galmod}
\begin{tabular}{lll}
\hline
Bulge & $(r_\mathrm{B}^2+r^2)^{-5/2}$ & $r_\mathrm{B}=0.38$ \\
\noalign{\vskip 0.1cm}
Disc & $\exp\left(-(\frac{R-R_\odot}{R_\mathrm{D}}+\frac{|z|}{z_\mathrm{D}})\right)$ & 
       $R_\odot=8.5\quad R_\mathrm{D}=3.5$\\
     & & $z_\mathrm{D}=0.2$ \\
\noalign{\vskip 0.1cm}
Halo & $(r_\mathrm{H}^{7/2}+(r^\prime)^{7/2})^{-1}$ &
       $ r_\mathrm{H}=1\quad q_\mathrm{H}=0.8$ \\
     & $r^\prime=(R^2+(z/q_\mathrm{H})^2)^{1/2}$ & \\
\hline
\end{tabular}
\end{table}

\subsection{The Galactic luminosity distribution model}

The model for the light distribution of the Galaxy consists of three spatial
components; a bulge with a Plummer law density distribution, a double
exponential disc and a flattened halo component for which the density falls of
as $r^{-3.5}$ at large distances from the Galactic centre.  The analytical
forms together with the scalelengths of these components are listed in
Table~\ref{tab:galmod}. The component luminosities are all normalized to their
luminosity density at the position of the Sun. We assume a total luminosity
density at the position of the Sun of $6.7\times10^{-2}$~L$_\odot$~pc$^{-3}$ and
bulge-to-disc and halo-to-disc luminosity density ratios of $5.5\times10^{-5}$
and $1.25\times10^{-3}$, respectively. This corresponds to a bulge-to-disc
luminosity ratio of $0.20$ and a halo-to-disc luminosity ratio of
$0.1667$. The corresponding total luminosity is $3.2\times10^{10}$~L$_\odot$.

The stars in this model are all assigned a spectral type and luminosity class
according to the Hess-diagram listed in Table 4--7 of \citet{MB81}. The
Hess-diagram provides the relative numbers of stars in bins of absolute
magnitude ($M_V$) and spectral type. These numbers integrated over spectral
type provide the luminosity function. We consider the Hess-diagram fixed for
all Galactic components and all positions throughout the Galaxy.

Finally, the kinematics are modelled assuming that each Galactic component
rotates with a constant velocity dispersion. For the bulge an isotropic
velocity dispersion is assumed. For the disc a different velocity ellipsoid is
assumed for each spectral type. For the halo the same velocity ellipsoid is
assumed regardless of spectral type. In all cases, the velocity ellipsoid is
assumed to be fixed in cylindrical coordinates. The velocity dispersions and
rotation velocities are listed in Table~\ref{tab:galkin}. The values for the
bulge have been taken from section 10.2.4 in \citet{BM98} (except for the
rotation which we have arbitrarily set to zero). The values for the disc are
from table 7--1 in \citet{MB81} (ignoring the vertex deviations of the
velocity ellipsoids) and for the halo table 10.6 in \citet{BM98} was used.

Although this is a highly simplified model of the Galaxy it is good enough for
providing the desired `background' distribution in phase-space.

\begin{table}
\caption{The kinematics of the three components of the model Galaxy. The
rotation velocity and velocity ellipsoid(s) of each component are listed in
units of km~s$^{-1}$. The velocity ellipsoid of the bulge is isotropic. For the disc
and halo the velocity ellipsoids are given in Galactocentric cylindrical
coordinates as function of stellar spectral type.}
\label{tab:galkin}
\begin{tabular}{llllll}
\hline
Component & $v_\mathrm{rot}$ & \multicolumn{4}{l}{Velocity ellipsoid} \\
\hline
Bulge & 0 & \multicolumn{4}{l}{$\sigma=110$} \\[2pt]
Disc  & 220 & Sp.\ Type & $\sigma_R$ & $\sigma_\theta$ &
$\sigma_z$ \\
      & & O & 10  & 9   & 6 \\
      & & B & 10  & 9   & 6 \\
      & & A & 20  & 9   & 9 \\
      & & F & 27  & 17  & 17 \\
      & & G & 32  & 17  & 15 \\
      & & K & 35  & 20  & 16 \\
      & & M & 31  & 23  & 16 \\[2pt]
Halo  & 35 &\multicolumn{4}{l}{$\sigma_R=135,\quad \sigma_\theta=105,\quad
\sigma_z=90$} \\
\hline
\end{tabular}
\end{table}

\subsection{Practical implementation of the Monte Carlo model}

A straightforward realization of our Galaxy model consists of independently
generating for each star a random space position, absolute magnitude and
population type, followed by the corresponding kinematics. However, the
\textit{Gaia} survey will be magnitude limited and in that case this simple
Monte Carlo procedure is potentially very wasteful. For each star one has to
check whether it will be bright enough in apparent magnitude to be included in
the survey. If the magnitude-limited volume is small compared to the volume
occupied by the Galaxy, a large fraction of stars will not appear in the final
survey. This is a particularly severe problem for intrinsically faint stars,
which are the most abundant.

The strategy we use minimises wasted effort by generating stars only within
the magnitude limited sphere. Let us assume a Galaxy whose density (in number
of stars per unit volume) is given by $\rho_\mathrm{G}(\mathbf{r})$ and whose
luminosity function (fraction of stars with absolute magnitudes between $M_V$
and $M_V+1$) is given by $\phi_\mathrm{G}(M_V)$, regardless of position in the
Galaxy. An observer at the position $r_\mathrm{obs}$ within the Galaxy
conducts a magnitude limited survey with a cut-off at $m_\mathrm{lim}$. This
means that for stars with absolute magnitudes between $M_V$ and $M_V+1$ the
number of stars in the survey will be:
\begin{eqnarray}
n^\mathrm{G}_\mathrm{vis}(M_V) & = & \int_{\Omega_V}
\rho_\mathrm{G}(\mathbf{r}) \phi_\mathrm{G}(M_V) d\mathbf{r} \\
 & = & \phi_\mathrm{G}(M_V) \int_{\Omega_V} \rho_\mathrm{G}(\mathbf{r})
d\mathbf{r} \,, \nonumber
\end{eqnarray}
where $\Omega_V$ is the magnitude limited sphere centred on the observer,
whose radius is given by the maximum distance at which the star is still
visible:
\begin{equation}
r_\mathrm{max}=10^{0.2(m_\mathrm{lim}-M_V)+1} \,.
\end{equation}
The total number of stars can be found by integrating over all absolute
magnitudes:
\begin{eqnarray}
N^\mathrm{G}_\mathrm{vis} & = & \int n^\mathrm{G}_\mathrm{vis}(M_V) dM_V
\nonumber \\
& = & \int \phi_\mathrm{G}(M_V) \left[ \int_{\Omega_V}
  \rho_\mathrm{G}(\mathbf{r}) d\mathbf{r} \right] dM_V\\
& = & \int_{M_V^1}^{M_V^2} \phi_\mathrm{G}(M_V) \zeta_\mathrm{G}(M_V) dM_V \,,
\nonumber
\end{eqnarray}
where $M_V^1$ and $M_V^2$ are the faint and bright limits of the luminosity
function and $\zeta_\mathrm{G}(M_V)$ is the integral within square brackets
which represents the number of stars in the Galaxy within the magnitude
limited volume $\Omega_V$. Taking the product of $\zeta_\mathrm{G}(M_V)$ with
$\phi_\mathrm{G}(M_V)$ brings down $\phi_\mathrm{G}(M_V)$ to the actual
number of stars of magnitude between $M_V$ and $M_V+1$ seen in the survey.

If we generate a Monte Carlo simulation of the magnitude limited survey with
$N_\mathrm{MC}^\mathrm{G}$ `stars' such that the luminosity function is
randomly sampled and the density of the galaxy is fully and randomly sampled
within the $\Omega_V$ corresponding to each $M_V$, the Monte Carlo simulation
will represent a `fair' sample of what would be obtained in the magnitude
limited survey.

\begin{figure}
\includegraphics[width=\columnwidth]{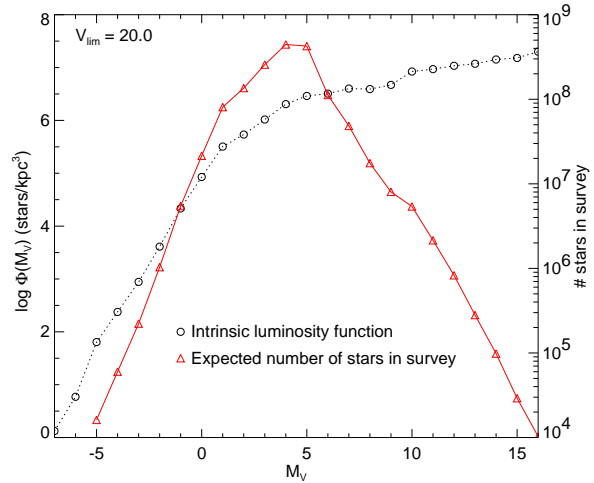}
\caption{Intrinsic and weighted luminosity functions $\phi_\mathrm{G}$ (left
  vertical scale) and $\psi_\mathrm{G}$ (right vertical scale) for our galaxy
  model. The circles indicate $\phi_\mathrm{G}$ and the triangles
  $\psi_\mathrm{G}$, plotted as the actual number of stars expected in the
  survey.}
\label{fig:lumfuncs}
\end{figure}

One way of ensuring this proper random sampling is as follows. We define a
weighted luminosity as:
\begin{equation}
\psi_\mathrm{G}(M_V) = \phi_\mathrm{G}(M_V) \zeta_\mathrm{G}(M_V) \,,
\end{equation}
and draw a random $M_V$ from $\psi_\mathrm{G}(M_V)$. Subsequently we draw a
random position from within the corresponding magnitude limited sphere
$\Omega_V$, using $\rho_\mathrm{G}$ as a probability function within this
volume\footnote{this means a renormalization such that $\int_{\Omega_V}
\rho_\mathrm{G}=1$}. We take an upper limit to the radius of the magnitude
limited volume of 100~kpc. In Fig.~\ref{fig:lumfuncs} we show both the
intrinsic and weighted luminosity functions $\phi_\mathrm{G}$ and
$\psi_\mathrm{G}$ for our galaxy model. The intrinsic luminosity function was
obtained by summing over spectral type in Table 4--7 of \cite{MB81} and the
weighted luminosity function was calculated for $-5\leq M_V\leq+16$.

The Monte Carlo realization of our galaxy model thus proceeds as
follows. Given that the luminosity function and spectral type distribution is
assumed the same for all Galactic components and positions within the Galaxy,
we start by drawing a random value for $M_V$ and spectral type from the Hess
diagram, weighted in luminosity according to the procedure outlined above. The
values for $M_V$ are only tabulated as integers which will produce
discontinuities in the the generated stellar positions. To avoid this we add a
random fraction between 0 and 1 to each generated value of $M_V$.

Once the absolute magnitude of a star is chosen, the Von Neumann rejection
technique \citep[e.g.][Section 7.3]{NRECIP} is used to draw a random 3D position
within the corresponding magnitude limited sphere. The sum of the densities of
each Galactic component, with the appropriate relative amplitudes, is used for
this task. Once the position is obtained, the Galactic component to which the
star belongs is assigned using the relative amplitude of each component at
that position. Finally, a velocity vector is randomly drawn from the
kinematical model appropriate for the Galactic component to which the star has
been assigned.

We use the Von Neumann rejection technique to generate the positions because
the total density distribution is a rather complicated function of the
cylindrical coordinates $R$ and $z$ when these are centred on the
observer. Although simple to implement, this technique turns out be very
inefficient for our case. We want to generate stellar positions within the
magnitude limited sphere $\Omega_V$ centred on the position of the Sun. To
apply the rejection technique we need a function which is everywhere within
$\Omega_V$ larger than $\rho_G$ \citep[see section 7.3 of][]{NRECIP}. The
simplest function to use is a constant equal to the maximum density within
this sphere, however its value cannot be obtained in a simple way. So one is
then forced to use the density at the Galactic centre as the upper bound to
$\rho_G$ and then first generate a position for the star in $\Omega_V$ before
applying the Von Neumann technique to decide on whether or not to retain the
star. For positions far away from the Galactic centre (either faint stars near
the Sun or brighter stars far out in the halo) this becomes very inefficient
owing to the large density contrast.

To prepare for the volume of data that \textit{Gaia} will deliver and in order to have
a realistic number of Galactic background stars with respect to the satellite
stars, we want to achieve a one-to-one sampling of the \textit{Gaia} survey in our
Monte Carlo model of the Galaxy. This implies that we need to generate the
positions of $\sim1.5\times10^9$ stars. This is a task that would take a
prohibitive amount of time, even using a fast cluster of Pentium CPUs as we
did (see Section~\ref{sec:nbody}).

In practice the tracing of satellites on the sky in the parts of the Galactic
plane towards the inner Galaxy will be very hard owing mainly to the large
extinction in those directions. Therefore we decided not to simulate the part
of the sky with Galactic longitude between $-90^\circ$ and $+90^\circ$, and
with Galactic latitude between $-5^\circ$ and $+5^\circ$. In our Galactic
model $\sim80$ per cent of the stars lie in this region of the sky as seen
from the Sun for a survey limited at $V=20$. Hence we now have to simulate
`only' $\sim 3\times10^8$ stars. Moreover, cutting out this part of the sky
(and thus the Galactic centre) allows us to use a much smaller upper bound for
the density which turns out to be the biggest contribution to improving
efficiency. Now the entire Monte Carlo Galaxy can be generated in two weeks'
time on a Beowulf computer.

The resulting simulated survey of the Galaxy is thus fully sampled and
contains $3.5\times10^8$ stars. For each of these we generated the 6
phase-space coordinates, an absolute magnitude, a population type (bulge, disc
or halo) and a spectral type. In Section~\ref{sec:modelgaia} we discuss in
detail how these data were converted into a simulated \textit{Gaia} survey
consisting of positions on the sky, parallaxes, proper motions, radial
velocities and photometric information.

\begin{figure}
\includegraphics[width=\columnwidth]{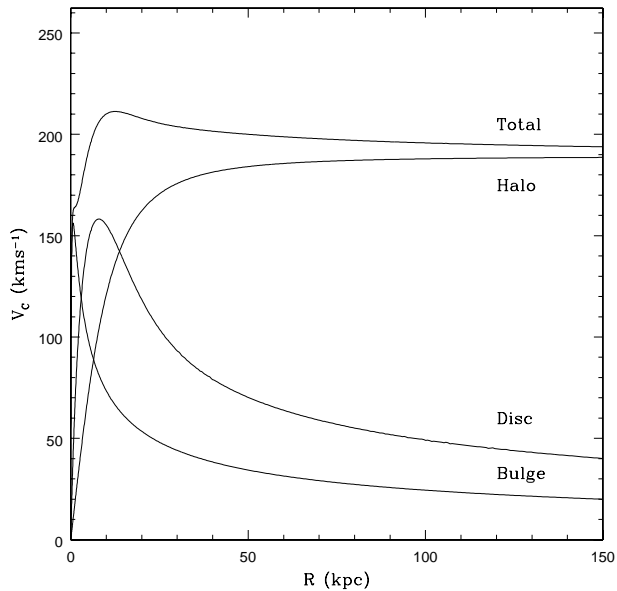}
\caption{Rotation curve for our Milky-Way mass model used in the $N$-body
simulations.}
\label{fig:galvc}
\end{figure}

%%%%%%%%%%%%%%%%%%%%%%%%%%%%%%%%%%%%%%%%%%%%%%%%%%%%%%%%%%%%%%%%%%%%%%%%%%%%%%%%%

\section{$N$-body Simulations}
\label{sec:nbody}
In this section we briefly describe the numerical technique used to study the
disruption of satellites by the Milky Way's gravitational potential. Given
that our aim is to identify the satellite remnants, we have chosen a rigid
potential to represent the Milky Way while self-consistent $N$-body
realizations are employed for the satellites.

\begin{table}
\caption{Parameters for our Milky Way mass model used in the $N$-body
simulations.}
\label{tab:galnbody}
\begin{tabular}{lll}
\hline
Disc & Bulge & Halo \\
\hline
$M_\mathrm{d} = 5.6\times 10^{10}\, \mathrm{M}_\odot$ & $M_\mathrm{b} =
1.4\times 10^{10} \, \mathrm{M}_\odot$ & $v_\mathrm{h} = 186$ kms$^{-1}$ \\
$R_\mathrm{d} = 3.5$ kpc & $a = 630$ pc & $R_\mathrm{C} = 12$ kpc \\
$\beta^{-1} = 700$ pc & & $q_\mathrm{\Phi_h} = 0.8$ \\
\hline
\end{tabular}
\end{table}

\subsection{Numerical models}
\subsubsection{Galaxy mass model}

Our Galaxy mass model for the $N$-body simulations is a composite consisting
of a disc, a bulge and a halo. The axisymmetric disc component is represented
by the following potential-density pair (Quinn \& Goodman 1986):

\begin{eqnarray}
\rho_\mathrm{d}(R,z) &=& {{M_\mathrm{d}\beta}\over{4\pi R_\mathrm{d}^2}}
      \exp(-R/R_\mathrm{d} -\beta|z|) \\
\Phi_\mathrm{d}(R,z) &=& -{{GM_\mathrm{d}}\over{R_\mathrm{d}^3}}
\int_\mathrm{0}^\mathrm{\infty} dk {{ k
\mathrm{J_o}(kR)}\over{(k^2+\beta^2)^{3/2}}}
{{\beta^2}\over{\beta^2-k^2}}
\nonumber \\
& {\phantom=} & \times [ {{\mathrm{e}^{-k|z|}}\over{k}} -
{{\mathrm{e}^{-\beta|z|}}\over{\beta}} ]
\nonumber
\end{eqnarray}
where $M_d$ is the mass of the disc, $R_d$ is the radial scalelength and
$\beta^{-1}$ is the vertical scaleheight.

A spherical Hernquist profile has been adopted for the bulge component
(Hernquist 1990):
\begin{equation}
\rho_b(r)={{M_b}\over{2\pi}}{{a}\over{r(r+a)^3}}
\end{equation}
where $M_b$ is the bulge mass and $a$ its scalelength. Another key ingredient
of our model is the massive dark halo which is represented by a logarithmic
potential \citep{BT87}.

\begin{equation}
\Phi_\mathrm{h} = v_\mathrm{h}^2\ln(R_\mathrm{C}^2+R^2+z^2/
q_\mathrm{\Phi_h}^2)
\end{equation}
where $v_\mathrm{h}$, $R_\mathrm{C}$ and $q_\mathrm{\Phi_h}\equiv
(c/a)_\mathrm{\Phi_h}$ are constants.  The last parameter is the flattening of
the halo potential.  Table~\ref{tab:galnbody} summarises the parameters of our
Milky Way mass model and its rotation curve is shown in
Figure~\ref{fig:galvc}. The total mass of the Milky Way model, truncated at
200~kpc, is $1.7\times10^{12}$~M$_\odot$.

\begin{table}
\caption{Self-consistent satellite models ($10^6$ particles).}
\label{tab:satnbody}
\begin{tabular}{lll}
\hline
&\multicolumn{1}{c}{S$_\mathrm{1}$} & \multicolumn{1}{c}{S$_\mathrm{2}$}\\[3pt]
\hline
$M_\mathrm{s}$  & $5.6\times 10^{7}\, \mathrm{M}_\odot$ & $2.8\times 10^{7}\, \mathrm{M}_\odot$ \\
$r_\mathrm{t}$  & $3150$ pc & $3150$ pc \\
$c           $  & $0.9$     & $0.9$     \\
\hline
\end{tabular}
\end{table}

Current models of large scale structure and galaxy formation postulate that
small objects are the first to form and then aggregate into larger systems
\citep[e.g.][]{KauffmannWhite1993}. In the non-baryonic CDM model, this
hierarchical assembly of haloes leads to triaxial shapes with average axis
ratios of $(c/a)_\mathrm{\rho_h}=0.5$ and $(b/a)_\mathrm{\rho_h}=0.7$
\citep{DubinskiC1991}. But, when a dissipative gas component is incorporated
they tend to be oblate with axis ratios of $(c/a)_\mathrm{\rho_h}=0.5$ and
$(b/a)_\mathrm{\rho_h}>0.7$ \citep{Dubinski1994}. Furthermore, haloes formed
within a $\Lambda$CDM model are more spherical with an axis ratio of
$(c/a)_\mathrm{\rho_h} \simeq 0.71$ while rounder haloes with
$(c/a)_\mathrm{\rho_h} \simeq 0.77$ are obtained for a $\Lambda$WDM model
\citep{Bullock2002}. The shape of our galaxy's halo is disputed and in this
respect the disruption of the Sagittarius dwarf galaxy provides a useful test
case. By studying tidal tails from this dwarf galaxy \citet{Ibata2001} ruled
out the possibility that the halo is significantly oblate (i.e. with an axis
ratio of $q_\mathrm{\rho_h} < 0.7$). However, \citet{MartinezDelgado2004}
suggest that the potential of the Milky Way dark halo, as constrained by
observations of the Sagittarius dwarf, is likely to be oblate with
$q_\mathrm{\Phi_h}=0.85$, while modelling of radial velocities of M-giants in
the Sagittarius dwarf by \citet{Helmi2004} suggests that the halo is prolate
with an average density axis ratio within the orbit of Sagittarius close to
$5/3$. Given all this evidence supporting a non-spherical halo, we have
adopted an oblate one to carry out our numerical simulations with an axis
ratio of $q_\mathrm{\Phi_h}=(c/a)_\mathrm{\Phi_h}=0.8$ which corresponds to a
flattening in density of $q_\mathrm{\rho_h}\approx 0.73$ and
$q_\mathrm{\rho_h}\approx 0.42$ in the inner and outer regions, respectively.

\begin{figure}
\includegraphics[width=\columnwidth]{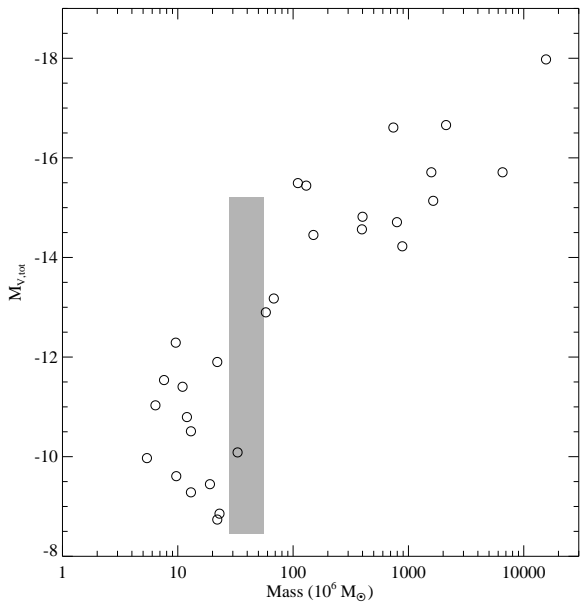}
\caption{Luminosity versus mass for dwarf galaxies of the Local Group (open
circles). The data were taken from \citet{Mateo1998} and the luminosities were
converted to total $V$-band magnitudes according to
$M_{V,\mathrm{tot}}=-2.5\log L_{V,\mathrm{tot}}+4.8$. The shaded region shows
the mass range of our $N$-body systems (masses of $2.8\times10^7$ and
$5.6\times10^7$~M$_\odot$) and the corresponding possible luminosity
range. The luminosity of a satellite of a given mass will depend on its mass
function and mass to light ratio.}
\label{fig:satlgroup}
\end{figure}

\subsubsection{Satellite mass models}

Positions and velocities for particles in our self-consistent satellite models
are randomly drawn from King models (King 1966). The latter are a series of
truncated isothermal spheres parametrized by a concentration defined as
$c=\log(r_\mathrm{t}/r_\mathrm{c})$ where $r_\mathrm{t}$ and $r_\mathrm{c}$
are the tidal and core radii, respectively. These models quite well fit the
dwarf spheroidal galaxies (Vader \& Chaboyer 1994). To completely define a
satellite we have to specify its mass and tidal radius in addition to its
concentration (see Table~\ref{tab:satnbody}). In all cases, each satellite
consists of $10^6$ equal-mass particles.

Figure~\ref{fig:satlgroup} shows the luminosity and mass distribution of real
Local Group dwarf galaxies. The masses and luminosities ($V$-band) vary
between $\sim10^7$ and $\sim10^{10}$~M$_\odot$, and $\sim3\times10^5$ and
$\sim10^9$~L$_\odot$, respectively \citep[see][]{Mateo1998}. This means that
our simulated dwarf galaxies fall in the low-mass end of the satellite mass
distribution. The grey shaded area in Fig.~\ref{fig:satlgroup} shows the range
of possible luminosities (given the observational data) for systems with
masses between $2.8\times10^7$ and $5.6\times10^7$~M$_\odot$. The range of
possible luminosities for our simulated dwarf galaxies is important for the
discussion in Section~\ref{sec:combined}.

\begin{table}
\caption{Orbital parameters for our experiments.}
\label{tab:satorb}
\begin{tabular}{ccccc}
\hline
Run  & Satellite & Pericentre & Apocentre &  Inclination    \\
     & Model           &   (kpc)    &    (kpc)  &  Angle          \\
\hline
1           & S$_\mathrm{1}$  & $8.75$ kpc  & $105$ kpc & $30^o$     \\
2           & S$_\mathrm{1}$  & $7$ kpc     & $60$ kpc  & $45^o$     \\
3           & S$_\mathrm{2}$  & $7$ kpc     & $80$ kpc  & $60^o$     \\
4           & S$_\mathrm{2}$  & $40$ kpc    & $60$ kpc  & $25^o$     \\
5           & S$_\mathrm{2}$  & $3.5$ kpc   & $55$ kpc  & $45^o$     \\
\hline

\end{tabular}
\end{table}

\begin{figure}
\includegraphics[width=\columnwidth]{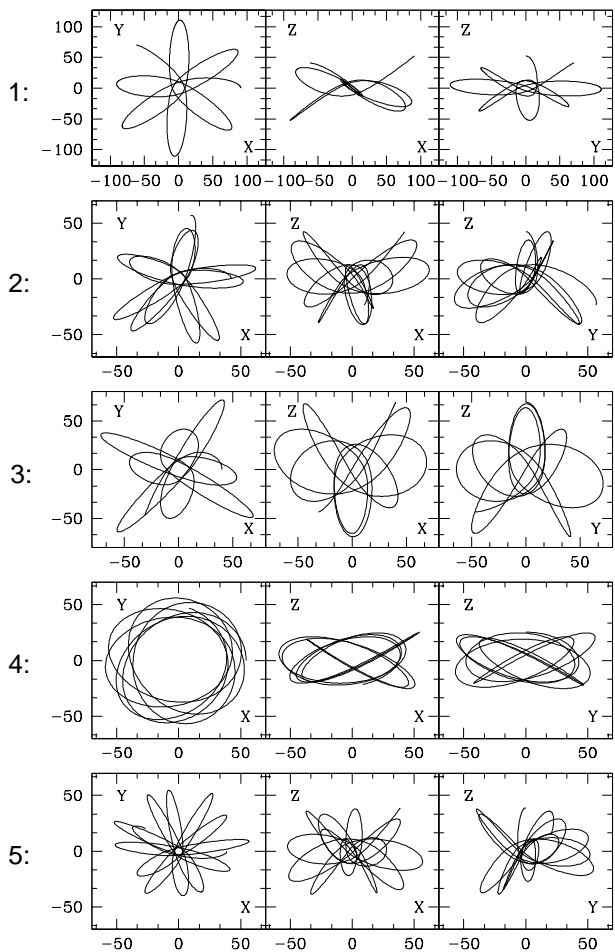}
\caption{Satellite orbits (see Table~\ref{tab:satorb}) projected onto the
principal axes of our Milky Way galaxy model. The Galactic disc corresponds to
the $XY$ projection. Distances are in kpc.}
\label{fig:allorb}
\end{figure}

\subsection{Numerical tools}

The simulations were evolved with a parallel tree-code for about $10$ Gyr.  It
uses a tolerance parameter of $\theta=0.75$ and a fixed integration time-step
of $1.3$~Myr \citep{Dubinski1996}. Forces between particles are computed
including the quadrupole terms and the gravitational potential uses a Plummer
softening length of $35$ pc. With these parameters the energy conservation in
all cases was better than $0.1$\%. All our numerical runs were performed on a
Beouwulf computer at the IAUNAM-Ensenada. It has $32$ Pentium processors
running at 450 MHz \citep{Velazquez2003}. Each simulation took 96 hours of
`wall-clock' time.

\begin{figure*}
\includegraphics[width=\textwidth]{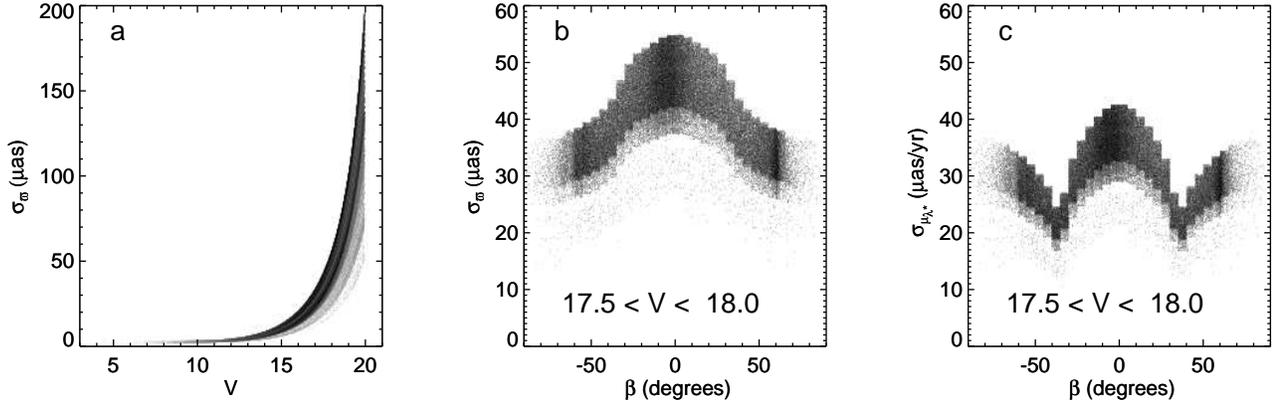}
\caption{The distribution of errors in the simulated \textit{Gaia} astrometric
  parameters for 1 million stars from the model Galaxy. The grey scale
  indicates the number of stars. Panel (a) shows the parallax errors as a
  function of $V$, the spread is caused by the colour dependence of the errors
  (see Appendix~\ref{ap:gaiamod}). Panel (b) shows the variation of the
  parallax errors with ecliptic latitude $\beta$ and panel (c) shows the
  dependence of the errors in $\mu_{\lambda*}$ on $\beta$. In order to clearly
  bring out the variation over the sky, the magnitude range of the stars was
  restricted for panels (b) and (c). The discrete steps in (b) and (c) are
  caused by the discrete tabulation of the dependence of the errors on
  $\beta$. There is no dependence of the errors on ecliptic longitude and
  $\sigma_{\mu_\beta}$ shows practically no variation with $\beta$.}
\label{fig:gaiaerrs}
\end{figure*}

\subsection{Numerical experiments}

A total of five orbits were chosen to follow up the disruption of the
satellite models. The parameters that define them are listed in
Table~\ref{tab:satorb}. The last column refers to the initial inclination
angle of the orbit with respect to the plane of the Galactic disc. Notice that
most of these orbits have a pericentre radius inside the Solar radius.
Figure~\ref{fig:allorb} shows the different projections of these orbits onto
the main axes of our Milky Way mass model where the plane of the Galactic disc
corresponds to the $XY$ projection. Since the Milky Way galaxy is represented
by a rigid potential and dynamical friction is ignored, the orbital decay of
the satellite is not taken into account.

%%%%%%%%%%%%%%%%%%%%%%%%%%%%%%%%%%%%%%%%%%%%%%%%%%%%%%%%%%%%%%%%%%%%%%%%%%%%%%%%%%

\section{Modelling the \textit{Gaia} Survey}
\label{sec:modelgaia}

In this section we describe how the stellar positions and velocities from the
Monte Carlo model of the Galaxy and the $N$-body models of the satellites are
converted to the data that \textit{Gaia} will deliver: sky-positions;
parallaxes; proper motions; and radial velocities. We also describe how to put
together the Galaxy and satellite simulations taking care to correctly account
for the variation of the number of visible stars along the orbit of the
satellite, while using as much of the $N$-body particles as possible.

\subsection{The Galactic data}
\label{sec:galdata}

To convert phase-space coordinates in the Galaxy model to \textit{Gaia} data
we first translate them to the barycentric coordinate frame, which has the
same axes as the Galactocentric system used for the simulations but with its
origin at the Sun's position and velocity. The barycentric position
$\vect{r}^\mathrm{b}=\vect{r}-\vect{r}_\odot$ and velocity
$\vect{v}^\mathrm{b}=\vect{v}-\vect{v}_\odot$ are converted to the five
astrometric parameters and the radial velocity $v_\mathrm{rad}$. The
astrometric observables are the Galactic coordinates $(\ell,b)$, the parallax
$\varpi$, and the proper motions $\mu_{\ell*}=\mu_\ell\cos b$ and $\mu_b$. The
units of parallax and proper motion are micro-arcsecond ($\mu$as) and
$\mu$as~yr$^{-1}$. The radial velocity is given in km~s$^{-1}$.

Subsequently the expected \textit{Gaia} errors are added to the observables
according to the accuracy assessment described in Chapt.\ 8 of the
\textit{Gaia} Concept and Technology Study Report \citep{ESA2000}. The
dominant variation of the size of the errors is with the apparent brightness
of the stars in addition to which there is a dependence on colour and position
on the sky. Thus, as a first step we calculate the $V$-band magnitude of each
star from its parallax and assign a $(V-I)$ colour based on the spectral type
of the star. The relation between spectral type and $(V-I)$ was taken from
\citet{Cox2000}. Note that no extinction is included in our model Galaxy.

Like Hipparcos \citep[see][]{ESA1997} \textit{Gaia} will be a continuously
scanning spacecraft and will accurately measure one-dimensional coordinates
along great circles in two fields of view simultaneously, separated by a
well-known angle. The accuracy of the astrometric measurements will thus
depend on how the parallactic plus proper motions of a particular object
project onto these great circles. Because \textit{Gaia} will collect its
measurements from an orbit in the plane of the Ecliptic the astrometric errors
will vary over the sky as a function of the Ecliptic latitude $\beta$. For the
positions and proper motions the latter variation of the precision is caused
mainly by the variation in the number of times an object is observed by
\textit{Gaia}, which is dependent on the details of the way \textit{Gaia}
scans the sky over the mission lifetime. For the parallax measurements the
dominant factor is the decrease in measurement precision from the ecliptic
pole to the equator owing to the way the parallax ellipse of a star projects
onto a great circle scan (the parallax factor). The precision of the radial
velocity measurements varies as a function of $V$ and spectral type but is
assumed not vary over the sky in our simulations.

The dependence of the errors on the Ecliptic coordinates $(\lambda,\beta)$ of
a star is taken into account in the simulations by rotating the astrometric
data to Ecliptic coordinates, adding the errors and rotating the result back
to the Galactic reference frame. As a consequence correlations will be
introduced between the errors in $\ell$ and $b$ as well as $\mu_{\ell*}$ and
$\mu_b$, even in the absence of any correlation in the Ecliptic reference
frame.

Figure~\ref{fig:gaiaerrs} illustrates the dependence of the errors in the
astrometric data on the apparent brightness of the star and its position on
the sky. The variation with stellar magnitude is the strongest. We show only
the variation of parallax error $\sigma_\varpi$ with $V$ as the variation is
similar for the other astrometric parameters. The spread in $\sigma_\varpi$ at
each value of $V$ is caused by the colour dependence of the astrometric errors
(smaller errors for redder stars, see Appendix~\ref{ap:gaiamod}). The
variation of the errors over the sky is shown in Ecliptic coordinates
$(\lambda,\beta)$ and is prominent only for $\sigma_\varpi$ and
$\sigma_{\mu_{\lambda*}}$. The errors only vary as a function of ecliptic
latitude $\beta$. Finally we point out that the relative error in the sky
positions is at most $\sim 10^{-9}$ (which is 200~$\mu$as expressed in
radians) meaning that distance errors perpendicular to the line of sight to
even the most distant stars (at about 100 kpc) amount to no more than about
$10^{-4}$ pc. Hence we will ignore errors on the sky positions in the rest of
the paper.

Finally, the steps taken to construct the simulated \textit{Gaia} catalogue are
summarized here:
\begin{enumerate}
\item[(1)] From \vect{r} and \vect{v} calculate the barycentric position and
velocity.
\item[(2)] Transform these to the astrometric parameters in the Galactic coordinate
  system and the radial velocity.
\item[(3)] Calculate $V$ from $M_V$ and $\varpi$.
\item[(4)] Find the $(V-I)$ colour corresponding to the spectral type of the star.
\item[(5)] Rotate the astrometric parameters to the Ecliptic coordinate system.
\item[(6)] Add the astrometric errors as a function of $V$, $(V-I)$ and
  Ecliptic latitude.
\item[(7)] Rotate the resulting astrometric parameters back to the Galactic
  coordinate system.
\item[(8)] Add radial velocity errors.
\end{enumerate}
For details please see Appendix~\ref{ap:gaiamod}.

\subsection{The satellite data}
\label{sec:satdata}

Given $V$ and $(V-I)$, the satellite stars are treated in exactly the same way
as the stars in the Monte Carlo Galaxy where the calculation of the
astrometric and radial velocity data is concerned. Hence here we describe only
how the absolute magnitudes and colours of the satellite stars are chosen.

We assume our simulated satellites represent disrupted dwarf galaxies and in
this and the following sub-section we will make a distinction between real
dwarf galaxies and our simulated $N$-body satellites. We will refer to the
former as `dwarf galaxies' or `satellites' and to the latter as `$N$-body
systems' or `simulated satellites'.

All $N$-body systems are assumed to represent a single stellar population of a
certain age and metallicity, which forms a single isochrone in a
colour-magnitude diagram. Combining this isochrone with a mass function
$\xi(m)$ for the satellite stars, we assign to each $N$-body particle a value
for $M_V$ and $(V-I)$. We use the isochrones by \citet{Girardi2000}. These
isochrones have been transformed to several different photometric systems as
described in \citet{Girardi2002}. The procedure for a particular star is to
first generate its mass and then to interpolate the isochrone in mass to find
the corresponding values of $M_V$ and $(V-I)$.

The colour-magnitude diagrams for a number of Local Group dwarf galaxies are
very similar to those of Galactic globular clusters \citep[see
e.g.,][]{Feltzing1999, Monkiewicz1999}. This suggests that we can use a
globular cluster like mass function for our $N$-body systems. Based on the
study of a dozen globular clusters \citet{Paresce2000} propose a lognormal
mass function for these systems for stars below about 1~M$_\odot$. This mass
function has a characteristic mass of $0.33$~M$_\odot$ and a standard
deviation of $0.34$. The slope of the mass function between $\sim0.3$ and
$\sim1$~M$_\odot$ is similar to that of a power-law mass function for which
$\xi(m)\propto m^{-\alpha}$, with $\alpha\approx1.5$. For N-body snapshots
older than 5~Gyr the stars in the simulated satellites will all have masses
less than about 2~M$_\odot$ and for simplicity we use a single power-law mass
function $\xi(m)\propto m^{-1.5}$. Hence we may be overestimating the number
of low-mass stars (below $0.3$~M$_\odot$) in our $N$-body systems with respect
to the more massive stars.

\subsection{Converting satellite $N$-body particles to stars}
\label{sec:combined}

Having made a considerable effort to realistically --- in terms of the number
of stars --- simulate the Galaxy as it will be seen by \textit{Gaia}, we want to ensure
that the satellite $N$-body particles are properly converted to simulated
stars representing a dwarf galaxy. Getting this right is not trivial and we
explain below why this is the case and describe in detail our solution to this
problem.

Given a certain distribution of stars along the orbit of a particular dwarf
galaxy (corresponding to an $N$-body `snap-shot'), the number of stars from
this satellite that will end up in the \textit{Gaia} catalogue depends on three
factors:
\begin{enumerate}
\item The overall number of stars (i.e., luminous particles) in the dwarf
  galaxy. This number is determined by its overall luminosity and the stellar
  mass function.
\item The \textit{Gaia} survey limit. This magnitude limit determines the faintest
  dwarf galaxy stars visible by \textit{Gaia}.
\item The variation of the number of visible satellite stars along the
  orbit. This variation is caused by the variation in distance from the dwarf
  galaxy stars to the Sun.
\end{enumerate}

If we take the $N$-body particles as representing a sampling of the
distribution of the real satellite stars along the orbit, then the expected
number of satellite stars visible in the \textit{Gaia} survey can be calculated as
follows. The overall number of stars (luminous particles) $N_\mathrm{lum}$ in
a real dwarf galaxy can be obtained from its total luminosity in the $V$-band
$L_{V,\mathrm{tot}}$ and its mass function $\xi(m)$:
\begin{equation}
\label{eq:deflumv}
N_\mathrm{lum}=\frac{L_{V,\mathrm{tot}}}{\langle L_V\rangle}\,, \quad
\langle L_V\rangle=\frac{\int L_V(m)\xi(m)dm}{\int \xi(m)dm}\,.
\end{equation}
$L_V(m)$ is the $V$-band luminosity of a dwarf galaxy star with mass $m$ and
$\langle L_V\rangle$ is the average stellar luminosity in the dwarf galaxy. At
each distance $s_i$ at which an $N$-body particle is located, the fraction of
visible satellite stars is set by the magnitude limit $M_{V,\mathrm{max}}$
which is given by:
\begin{equation}
M_{V,\mathrm{max}}(s_i)=V_\mathrm{lim}-5\log s_i+5\,,
\end{equation}
where $V_\mathrm{lim}$ is the \textit{Gaia} survey limit. The fraction of visible
satellite stars $f_i$ at distance $s_i$ is then:
\begin{equation}
f_i(s_i)=\frac{\int_{m(M_{V,\mathrm{max}})}^{m_\mathrm{up}}\xi(m)dm}{\int
  \xi(m)dm}\,,
\end{equation}
where $m_\mathrm{up}$ is the upper mass limit of the mass function. The
overall fraction $f$ of visible satellite stars is then given by:
\begin{equation}
\label{eq:defoverallfrac}
f=\frac{\sum_i f_i(s_i)}{N_\mathrm{sim}}\,,
\end{equation}
where $N_\mathrm{sim}$ is now the total number of {\em simulated} satellite
particles. Hence the number of satellite stars expected in the \textit{Gaia}
survey is $fN_\mathrm{lum}$. Note that we assume that the mass function does
not vary along the orbit of the satellite. This may not be realistic if the
satellite preferentially looses lower mass stars owing to tidal forces.

\begin{figure}
\includegraphics[width=\columnwidth]{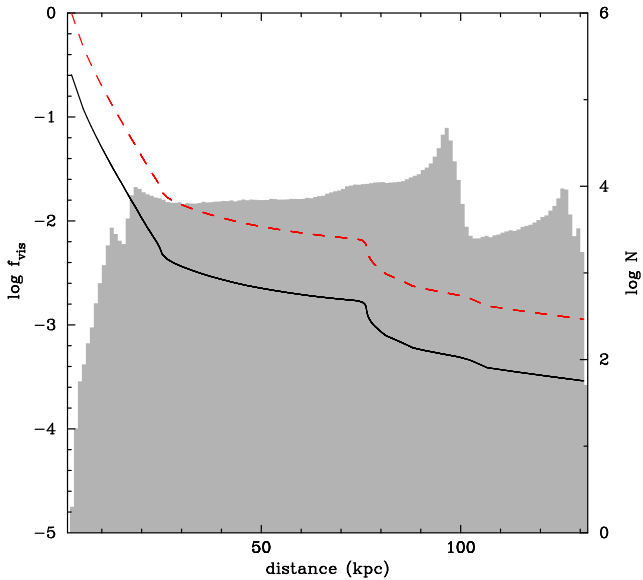}
\caption{Visible fraction of the simulated satellite stars at each $N$-body
  particle position as a function of the distance to the observer for $N$-body
  run no.\ 1. The shaded histogram (vertical scale on the right) shows the
  distribution of distances $s_i$. The solid line (vertical scale on the left)
  shows for each distance $s_i$ the fraction of visible stars $f_i$ calculated
  as in Section~\ref{sec:combined}. The features in this line reflect features
  in the luminosity function. The dashed line shows the fraction of visible
  stars when taking into account only the stars brighter than the faintest
  satellite star that enters the \textit{Gaia} survey. The fractions of visible stars
  were calculated using a 10 Gyr isochrone of metallicity $Z=0.004$ and a mass
  function $\xi(m)\propto m^{-1.5}$. The overall fraction of visible $N$-body
  particles is only $2.5\times10^{-3}$ and $10^{-2}$ for the two cases
  considered.}
\label{fig:visfrac1}
\end{figure}

Now, for a normal mass function the number of low-mass stars will heavily
dominate, which means that the fractions $f_i$ can be very small depending on
the distance $s_i$. Figure~\ref{fig:visfrac1} shows an example of this for a
snapshot at about 10 Gyr of $N$-body run no.\ 1 (see
Table~\ref{tab:satorb}). The stellar population of the simulated satellite is
represented by a 10~Gyr isochrone for a metallicity $Z=0.004$, with a mass
function $\xi(m)\propto m^{-1.5}$. Three things are shown as a function of the
distance $s_i$ of each $N$-body particle $i$. The shaded histogram shows the
distribution of distances $s_i$ and illustrates the wide variety of distances
at which the simulated satellite particles are located. The solid line shows
for each distance $s_i$ the fraction of visible stars $f_i$ calculated as
outlined above. The $N$-body particles are taken to represent a fully
populated mass function which in this case ranges between $0.15$ and
$0.88$~M$_\odot$. Note that $f_i$ is nowhere larger than $\sim 0.1$, while the
overall fraction $f$ of visible stars is $2.5\times10^{-3}$. This means that
if we assume that our $N$-body particles represent stars over the full range
of the satellite mass function, only $2\,500$ out of the 1 million simulated
stars will end up in our synthetic \textit{Gaia} catalogue!

\begin{figure}
\includegraphics[width=\columnwidth]{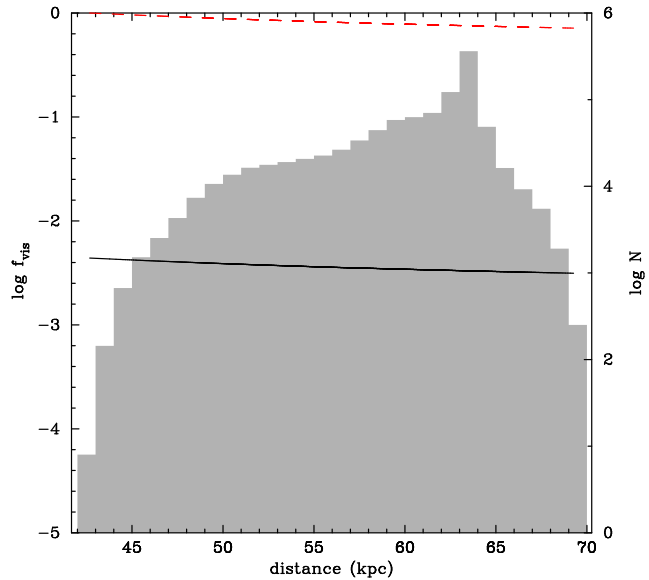}
\caption{Same as Fig.~\ref{fig:visfrac1} for the simulated satellite from
  $N$-body run no.\ 4. A 5 Gyr isochrone and the same mass function were
  used. Note the much smaller range in distances for this $N$-body system,
  resulting in a much larger fraction of visible stars if the luminosity
  function is cut off at $M_V\leq M_{V,\mathrm{faintest}}$. The overall
  fraction of visible stars is $3.4\times10^{-3}$ when using the full mass
  function and $0.78$ when only stars with $M_V\le M_{V,\mathrm{faintest}}$
  are considered. }
\label{fig:visfrac2}
\end{figure}

This is clearly a very wasteful way to make use of the (computationally)
expensive $N$-body calculations and the question is how to retain more
$N$-body particles in the synthetic \textit{Gaia} catalogue whilst at the same time
preserving the variation of the visible fraction of stars along the $N$-body
system's orbit. One straightforward improvement can be made by realizing that
for mass functions that extend all the way to late M-dwarfs most of the
generated synthetic satellite stars are too faint to ever make it into the
\textit{Gaia} survey, regardless of the position at which they are located along the
satellite orbit. To eliminate this problem we can first find the position in
the $N$-body snapshot that is closest to the Sun (at a distance
$s_\mathrm{min}$) and calculate the corresponding absolute magnitude,
$M_{V,\mathrm{faintest}}$, of the faintest satellite stars visible by
\textit{Gaia}. Subsequently in assigning luminosities to the simulated satellite stars
we use the luminosity function only for $M_V\leq M_{V,\mathrm{faintest}}$. The
result is shown in Fig.~\ref{fig:visfrac1} by the dashed line. The visible
fraction of stars increases everywhere by more than an order of magnitude due
to the exclusion of the faintest (never visible) stars. However the overall
fraction of visible stars is still very low at $9.8\times10^{-3}$. This is a
consequence of the large variation in distances for the orbit of the simulated
satellite from $N$-body run no.\ 1. Figure~\ref{fig:visfrac2} shows the
distribution of distances and values of $f_i$ for the satellite from run no.\
4 at 5 Gyr (using a 5 Gyr isochrone and the same mass function), which has a
much more circular orbit with much less distance variation (compare the orbits
in Fig.~\ref{fig:allorb}). For this $N$-body system one can retain about 78
per cent of the $N$-body particles if the luminosity function cut-off at
$M_V\leq M_{V,\mathrm{faintest}}$ is used.

A further improvement is now obvious. Instead of cutting off the luminosity
function at $M_{V,\mathrm{faintest}}$ we choose a brighter limit and retain
more of the $N$-body particles in the synthetic \textit{Gaia} catalogue. This
can be interpreted as simulating the fact that in practice one will use bright
`tracer' stars (such as those at the tip of the AGB or RGB) to map out the
satellite in phase-space. That is, we assume that the $N$-body particles only
represent stars at the high mass end of the satellite's mass function. In fact
most of the present surveys for substructure in the Galactic halo rely on
using tracer populations that are more easily isolated photometrically in
colour-magnitude diagrams \citep[see
e.g.,][]{Majewski2003,Martin04,RochaPinto2004,Helmi2004}.

In principle the luminosity function cut-off magnitude $M_{V,\mathrm{tracer}}$
can be chosen bright enough to retain all $N$-body particles. However, there
are two constraints on its value, the first of which is the obvious limit set
by the magnitude of the brightest star in the luminosity function. The second
constraint is set by the desire for a realistic simulation of a satellite as
it will appear in the \textit{Gaia} catalogue. By assuming that the $N$-body particles
represent only a tracer population we are effectively simulating a more
massive dwarf galaxy when converting the $N$-body results to \textit{Gaia}
observations. In reality we have simulated rather small dwarf galaxies (see
Fig.~\ref{fig:satlgroup}) which cannot be scaled up trivially to more massive
systems because the latter respond differently to the Galactic tidal
forces. Thus a balancing between the number of retained $N$-body particles and
the total mass (or luminosity) of a real dwarf galaxy that we wish to simulate
is required.

\begin{figure}
\includegraphics[width=\columnwidth]{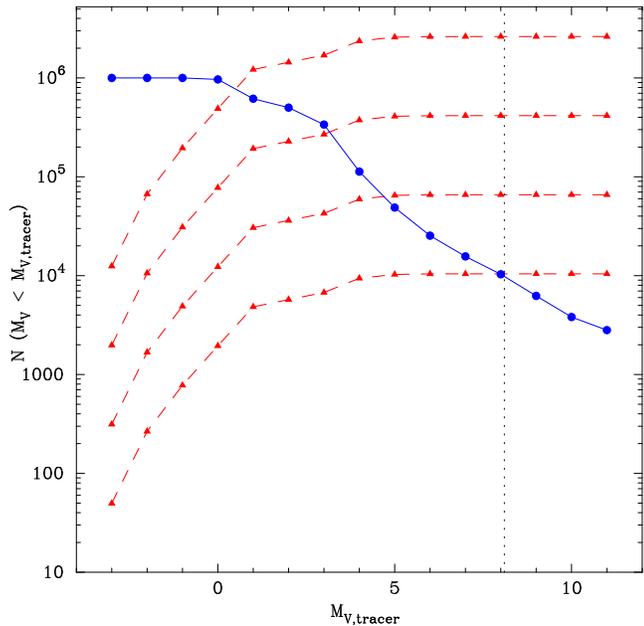}
\caption{Visible number of stars as a function of the luminosity function
  cut-off $M_{V,\mathrm{tracer}}$ for a satellite with its stars distributed
  along the orbit of $N$-body run no.\ 1. The dashed curves with triangles
  show the numbers for `real' dwarf galaxies (i.e., with fully populated
  luminosity functions) of total luminosities: $M_{V,\mathrm{tot}}=-17, -15,
  -13, -11$ (where $M_{V,\mathrm{tot}}=-2.5\log L_{V,\mathrm{tot}}+4.8$). The
  dotted line shows the value of $M_{V,\mathrm{faintest}}$ beyond which the
  dashed curves become flat. The solid curve with dots shows the numbers of
  visible stars for the $N$-body system, where now all stars are assumed to be
  brighter than $M_{V,\mathrm{tracer}}$. The fractions of visible stars were
  calculated using a 10 Gyr isochrone of metallicity $Z=0.004$ and a mass
  function $\xi(m)\propto m^{-1.5}$.}
\label{fig:sampling1}
\end{figure}

\begin{figure}
\includegraphics[width=\columnwidth]{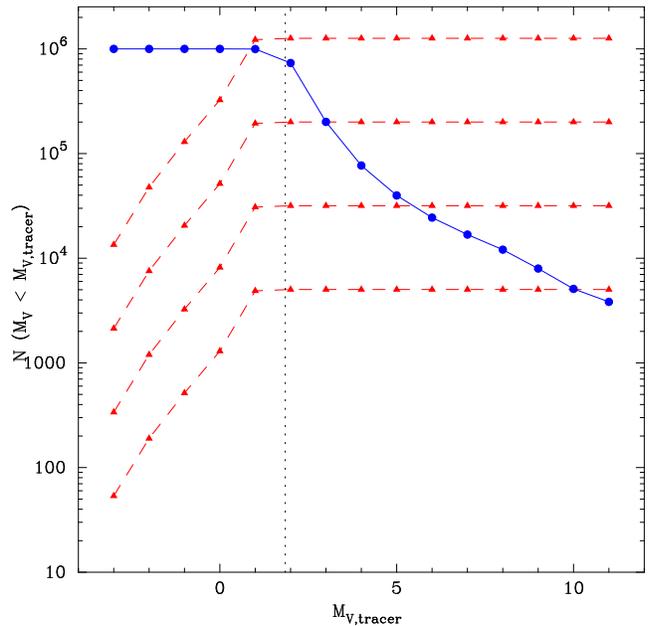}
\caption{Same as Fig.~\ref{fig:sampling1} for a satellite with its stars
  distributed along the orbit of from $N$-body run no.\ 4. A 5 Gyr isochrone
  and the same mass function were used. Note the brighter value of
  $M_{V,\mathrm{faintest}}$ which reflects the larger average distance of this
  satellite.}
\label{fig:sampling2}
\end{figure}

This balancing is illustrated in Figs.~\ref{fig:sampling1} and
\ref{fig:sampling2} for the $N$-body systems discussed above. The figures show
for the $N$-body system and for the corresponding real dwarf galaxies how the
number of stars visible by \textit{Gaia} changes as the magnitude cut-off of
the luminosity function, $M_{V,\mathrm{tracer}}$, changes. These numbers are
calculated according to equations
(\ref{eq:deflumv})--(\ref{eq:defoverallfrac}). For a real dwarf galaxy the
number of visible stars will increase as the value of $M_{V,\mathrm{tracer}}$
goes up, reflecting a probing of the luminosity function to fainter
magnitudes. At some point $M_{V,\mathrm{tracer}}$ becomes larger than
$M_{V,\mathrm{faintest}}$ and the number of visible stars stays constant
(probing deeper into the luminosity function only adds stars invisible to
\textit{Gaia}). This is shown by the dashed curves and triangles, where each
curve represents a dwarf galaxy of different $L_{V,\mathrm{tot}}$ (or total
mass). The dotted line shows the value of $M_{V,\mathrm{faintest}}$ beyond
which the curves are flat.

For the $N$-body system the number of visible particles will decrease as the
value of $M_{V,\mathrm{tracer}}$ goes up. This is because now {\em all}
particles are assumed to be brighter than $M_{V,\mathrm{tracer}}$ which leads
to a larger fraction of simulated stars that are fainter than the survey limit
as the luminosity function cut-off becomes fainter. In
Figs.~\ref{fig:sampling1} and \ref{fig:sampling2} the solid curve with dots
represents the visible number of stars for the $N$-body system. Note that this
curve continues to drop beyond $M_{V,\mathrm{tracer}}=M_{V,\mathrm{faintest}}$
which reflects the increasing number of simulated stars that will never enter
the \textit{Gaia} survey.

The balancing of the number of retained $N$-body particles and the total mass
of the dwarf galaxy we wish to simulate thus consists of choosing a dwarf
galaxy luminosity and then finding the value of $M_{V,\mathrm{tracer}}$ for
which the number of expected tracer stars from this satellite in the
\textit{Gaia} survey is equal to the number of retained $N$-body
particles. For $N$-body run no.\ 1 one can simulate an
$M_{V,\mathrm{tot}}=-17$ satellite by choosing
$M_{V,\mathrm{tracer}}\approx0.5$, which then results in between $7\times10^5$
and $8\times10^5$ $N$-body particles entering the synthetic \textit{Gaia}
catalogue. Note that in practice one can choose a value of
$M_{V,\mathrm{tracer}}$ corresponding to an approximate value of
$M_{V,\mathrm{tot}}$ by inspecting diagrams such as Figs.~\ref{fig:sampling1}
and \ref{fig:sampling2}, and then calculate the actual value of
$M_{V,\mathrm{tot}}$ based on the amount of $N$-body (tracer) particles that
enter the simulated \textit{Gaia} survey.

We now return to Fig.~\ref{fig:satlgroup} which shows that for the mass range
of the satellites we have simulated the total $V$-band magnitude can be at
most $M_{V,\mathrm{tot}}\approx-15$. From Figs.~\ref{fig:sampling1} and
\ref{fig:sampling2} one can see that this luminosity is simulated for
$M_{V,\mathrm{tracer}}\approx3$ and that then we retain about $2\times10^5$
$N$-body particles for each satellite. We still consider this wasteful and
decided to use $M_{V,\mathrm{tracer}}=0.5$ in order to retain many more
$N$-body particles. This implies an unrealistic value of $M_V\approx-17$ for
the absolute magnitude of the simulated satellites in the $V$-band. The main
problem being that the corresponding masses of more than $10^9$~M$_\odot$ (see
Fig.~\ref{fig:satlgroup}) would require simulations with at least an order of
magnitude more particles and with the effects of dynamical friction
included. This was beyond our computational capabilities at the time the
simulations were done. However, the purpose of this paper is not to provide
detailed and completely realistic simulations of the Galaxy and its
satellites. The main purpose is to get an insight into the challenges one will
have to deal with when searching for the debris of past merger events in the
\textit{Gaia} data. Therefore we feel it is more important to get a good sampling of
the distribution of satellite stars in the \textit{Gaia} catalogue, by artificially
increasing their numbers, than to construct a completely realistic simulation
of the satellite debris.

\begin{figure}
\includegraphics[height=\columnwidth,angle=90,clip=]{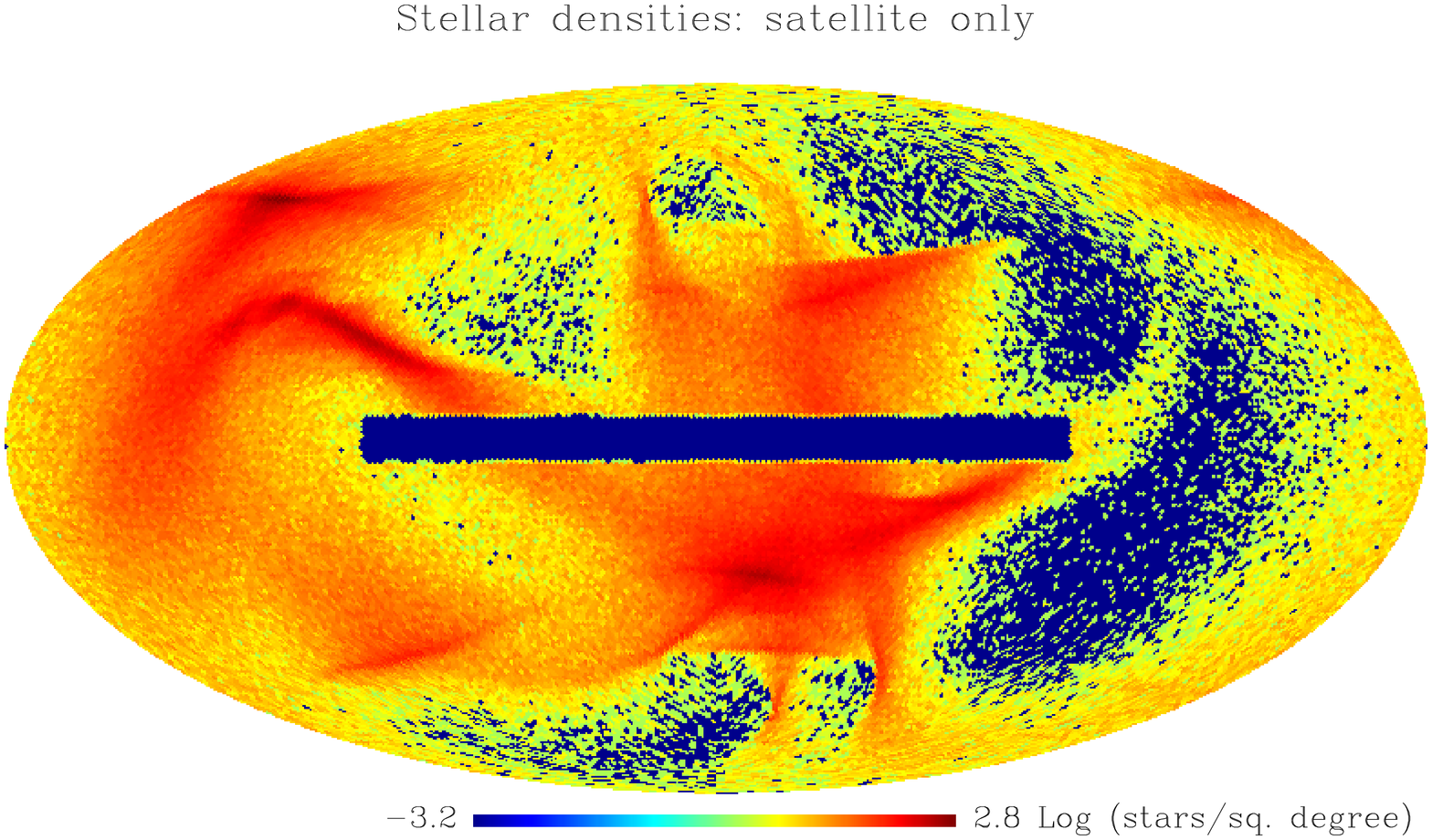}\\[5pt]
\includegraphics[height=\columnwidth,angle=90,clip=]{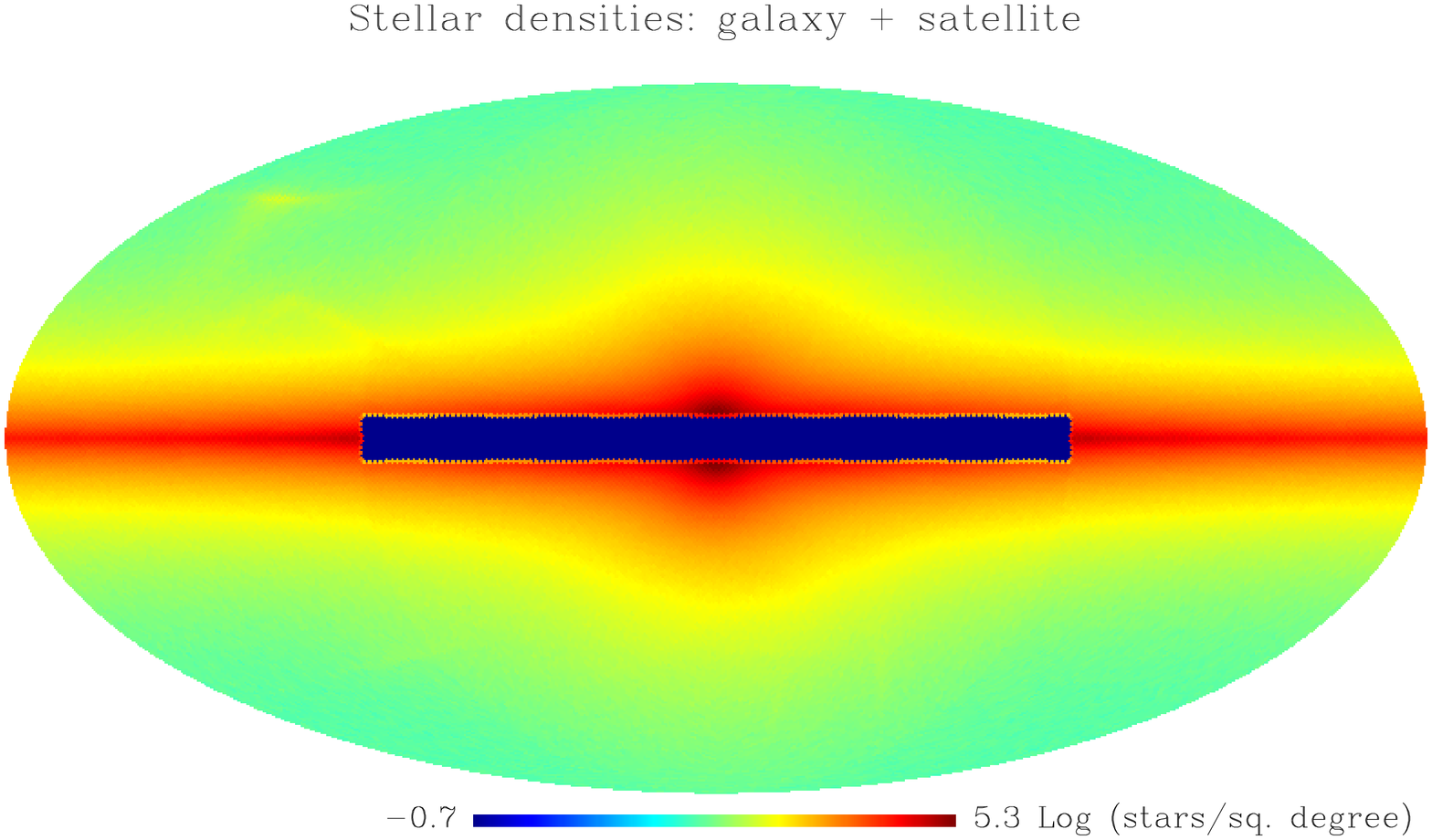}
\caption{Mollweide projections of the distribution on the sky of the stars in
  our simulated \textit{Gaia} catalogue. The top image shows only the
  satellites projected on the sky (from simulation runs no.\ 1 and 5 at
  10~Gyr, where an extra satellite was generated by reversing the angular
  momentum for run 1). The bottom image shows the Galaxy and satellite stars
  combined. Note how the satellite stars become almost totally invisible. The
  only conspicuous over-density is in the top left corner of the image. The
  dark rectangle is the area on the sky with coordinates
  $-90^\circ\leq\ell\leq +90^\circ$ and $-5^\circ\leq b\leq +5^\circ$ that was
  left out of the simulations. The figures were generated using the Healpix
  \citep*{Gorski1999} division of the celestial sphere into cells of equal
  area and counting the stars inside each cell. Dark areas correspond to empty
  cells while the individual dots indicate cells with 1 or more stars. Areas
  where all contiguous cells are filled appear as a grey scale.}
\label{fig:skyproject}
\end{figure}

Finally, we summarise here the steps for simulating the satellite galaxy as it
will appear in the \textit{Gaia} catalogue:
\begin{enumerate}
\item[(1)] calculate the distance to each $N$-body particle.
\item[(2)] Choose a value $M_{V,\mathrm{tracer}}$ for the minimum brightness of the
  tracer stars represented by the $N$-body particles.
\item[(3)] Re-normalise the luminosity function between $M_V(m_\mathrm{up})$ and
  $M_{V,\mathrm{tracer}}$ (i.e., all $N$-body particles represent stars
  brighter than $M_{V,\mathrm{tracer}}$).
\item[(4)] Generate simulated stars for each $N$-body particle position (distance)
  according to the renormalized luminosity function and decided whether or not
  they are bright enough to enter the \textit{Gaia} survey.
\item[(5)] For stars bright enough to enter the \textit{Gaia} survey assign a $(V-I)$ colour
  by interpolating in mass along the isochrone.
\item[(6)] Generate the \textit{Gaia} data according to the steps outlined in
  Section~\ref{sec:galdata}.
\end{enumerate}
This procedure ensures a properly sampled luminosity function along the orbit
of the dwarf galaxy.

Our procedure assumes that one can select the bright tracer stars a priori
which may not always be possible in practice and may not be desirable. A way
to get around this problem and still use as many $N$-body particles as
possible is the following. Assume that all $N$-body particles are brighter
than $M_{V,\mathrm{faintest}}$ and calculate the predicted number of $N$-body
particles that make it into the survey as for Figs.~\ref{fig:sampling1} and
\ref{fig:sampling2}. Calculate the ratio
$R=N_\mathrm{sim}/N_\mathrm{vis}$, where $N_\mathrm{vis}$ is the number of
visible particles for
$M_{V,\mathrm{tracer}}=M_{V,\mathrm{faintest}}$. Subsequently repeat steps
(1)--(6) above $R$ times for each $N$-body particle. That is, generate $R$
stars for each $N$-body particle and retain those that are bright enough to
enter the survey. This way all $10^6$ $N$-body particles can be retained in
the simulated \textit{Gaia} data whilst still retaining a proper luminosity function
sampling along the orbit of the dwarf galaxy. However one then faces the
problem of possibly having simulated a system that is too massive (but this
can be tuned by not retaining all $10^6$ particles) and of having simulated
$R$ stars at exactly the same position with the same velocity vector. It is
not clear how one would redistribute these extra stars along the dwarf galaxy
orbit such that their energy and angular momentum are consistent with that of
the $N$-body particles. We therefore decided not to pursue this option further.

To close this section we show in Fig.~\ref{fig:skyproject} the sky
distribution of the stars in our simulated \textit{Gaia} catalogue. The simulated
satellites included in these figures correspond to $N$-body runs 1 and 5 of
Table~\ref{tab:satorb} where a third satellite was generated by reversing the
angular momentum of the satellite from run 1. The main thing to note is how
the satellite stars are completely swamped by the Galactic stars in these sky
projections. This is not a new fact but serves to illustrate the need for
clever search techniques to find the debris of disrupted satellites, an issue
to which we turn in the next section.

%%%%%%%%%%%%%%%%%%%%%%%%%%%%%%%%%%%%%%%%%%%%%%%%%%%%%%%%%%%%%%%%%%%%%%%%%%%%%%%

\section{Results from \textit{Gaia} Survey Modelling}
\label{sec:results}

\citet{Helmi1999} showed that it will be quite difficult to find spatial
correlations in the debris originating from satellites, whose disruption
occurred in the early evolution of the Milky Way after several passages
through pericentre. In this respect, techniques to identify streamers like the
`Great Circle Cell Counts' proposed by \citet{Johnston1996} will be of little
practical use. This method requires that the orbital plane of the satellite
remains almost unchanged during its disruption process, so that the debris
occupies great circles on the sky as seen from the Galactic Centre. This only
occurs if the outer halo is spherical or if the satellite follows a polar or
planar orbit in the case of an oblate halo.
 
\citet{HWZZ1999} showed that useful information can be obtained from the
phase-space structure of these fossil remnants which allows tracing the merger
history of our Galaxy. Initially satellites are clumped in both the
configuration and velocity space, and also in the integrals of motion
space. The method of searching for debris in the integrals of motion space
relies on the basic assumption that these quantities are conserved, or evolve
only slightly, and hence their lumpiness should be preserved. For a
spherically symmetric halo the integrals of motions are the energy, total
angular momentum and angular momentum around the Galactic $Z$-axis,
$(E,L,L_z)$; for an axisymmetric halo $E$ and $L_z$ are preserved while $L$
evolves owing to the orbital precessing of the satellite but it still exhibits
some degree of coherence. The integrals of motion technique was applied by
\citet{Helmi2000} to an artificial \textit{Gaia} halo catalogue generated from
$N$-body simulations of the disruption of satellites in a Galactic
potential. From this study, they conclude that identifying substructure within
the stellar halo resulting from merger events in the past will be relatively
easy even after \textit{Gaia} observational uncertainties are taken into
account.

\begin{figure}
\includegraphics[width=\columnwidth]{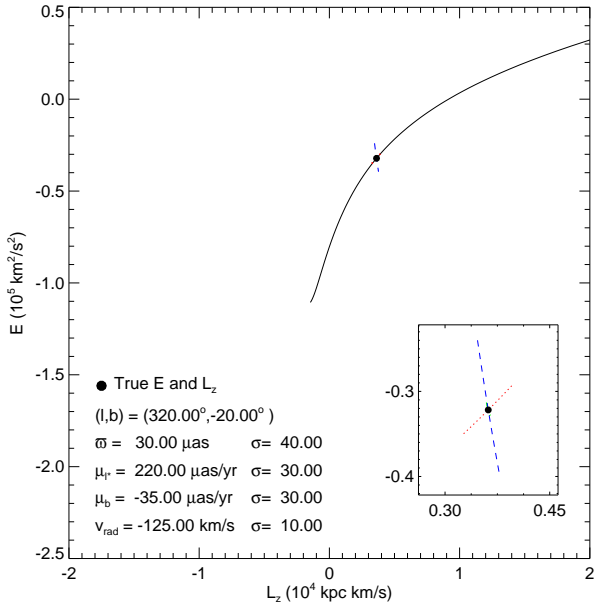}\\
\caption{Error propagation in the $E$--$L_z$ plane. The effect of the
  astrometric and radial velocity errors on the observed value of $E$ and
  $L_z$ is shown using the example of a particular star with \vect{r} and
  \vect{v} corresponding to the satellite from simulation run 1 at 10
  Gyr. Each line indicates how $E$ and $L_z$ change from the true value
  (indicated by the dot) as the observables are varied within $\pm3\sigma$
  from their true values. The values of the astrometric parameters, the radial
  velocity and the corresponding errors are listed in the legend. The solid
  line shows the effect of the error in the parallax. The dotted line shows
  the effect of errors in $\mu_{\ell*}$. The dashed line indicates the effect
  of radial velocity errors. The inset enlarges the region around the true
  value of $(E,L_z)$ in order to better show the effects of the proper motion
  and radial velocity errors. Note that the error in $\mu_b$ has almost no
  effect on $(E,L_z)$ for this particular star. The scale of the figure is the
  same as for Fig.~\ref{fig:elzspace}. For convenience the potential energy
  was computed using the potential from equations (8)--(10) in
  \citet{Helmi2000}.}
\label{fig:errorpropag}
\end{figure}

However \citet{Helmi2000} made two important simplifying assumptions. They
constructed a synthetic \textit{Gaia} catalogue consisting of stars that are
intrinsically as well as apparently bright ($M_V=1$, $V\leq15$), and which are
all located relatively near the Sun ($d<6$~kpc). The limit on apparent
brightness was set by the demand of having good radial velocities available
for their sample. Secondly, their simulated halo consists exclusively of
in-falling satellite Galaxies. These two assumptions lead to a sample of halo
stars which have small observational errors and will be highly clumped in the
integrals of motion space, making the disentangling of the different debris
streams relatively easy.

In reality there may well be a `background' population of halo stars that have
a smooth distribution in $(E,L,L_z)$ space owing to previous phase mixing and
dynamical friction or because they constitute a component that was formed in a
monolithic collapse and thoroughly mixed by violent relaxation. In addition
real debris streams will have a spread in stellar luminosities (even if only
bright tracers are selected such as in our simulations) and may be located
much further away. These will lead to larger observational errors which will
propagate into $(E,L,L_z)$ space. Both effects will smear out the signatures
of individual debris streams. We demonstrate this below for our simulations by
constructing $E$--$L_z$ diagrams. However, we will first discuss how the errors
in the astrometric parameters and the radial velocity propagate in the
$E$--$L_z$ diagrams.

\subsection{Error propagation in $E$--$L_z$ space}
\label{sec:elzerrors}

Figure~\ref{fig:errorpropag} schematically illustrates how the \textit{Gaia}
observational errors propagate into the $E$--$L_z$ plane. As an example a star
from the satellite of simulation run no.\ 1 is chosen and the parallax, proper
motions and radial velocities are varied by $\pm3\sigma$ from their true
values. The lines in the diagram trace the effect of varying each of the
observables individually. The measured parallax can in principle become
negative but then the calculation of the observed angular momentum becomes
ill-defined. Hence Fig.~\ref{fig:errorpropag} only shows the effect of the
positive branch of the measured parallaxes. The parallax error is by far the
dominant contribution to the errors in $E$--$L_z$ space and the observed values
of $(E,L_z)$ are smeared in a preferred direction. In addition it can be seen
that the angular momentum can change its sign for large parallax errors. This
happens because the star, which is located beyond the Solar circle in the
direction of $\ell=320^\circ$, can get placed inside the Solar circle if the
measured parallax is much larger than the real parallax (measured distance too
small). This puts the star at an observed position on the opposite side of the
Galactic centre which for an unchanged direction of its observed velocity
leads to a reversed angular momentum.

Another effect is caused by the non-linear dependence of $E$ and $L_z$ on
parallax, i.e.\ both quantities depend on functions of $1/\varpi$
(distance). This leads to systematic errors in the sense that the expectation
value for the observed $E$ or $L_z$ does not equal the true values of these
quantities. This is a well known problem when dealing with the determination
of distances to stars using parallaxes with relatively large errors \cite[see
e.g.,][]{Brown1997,Schroder2004}. The detailed systematic effects in the
$E$--$L_z$ plane depend on the actual distribution of the observed stars in
space, on the distribution of the parallax errors and, importantly, on the way
the sample of stars is selected. Any selection made on the size of the
parallax or on its quality ($\sigma_\varpi/\varpi$) will skew the underlying
distribution of true distances which will be reflected in the calculated
integrals of motion. A detailed investigation of these systematics is beyond
the scope of this paper but is important in the future use of \textit{Gaia} data to
study the integrals of motion in the Galaxy. Appendix~\ref{ap:calclz} contains
the equations for calculating the angular momentum of a star from astrometric
and radial velocity data. They can be used to further study the effects of
measurement errors.

\begin{figure*}
\includegraphics[width=\textwidth]{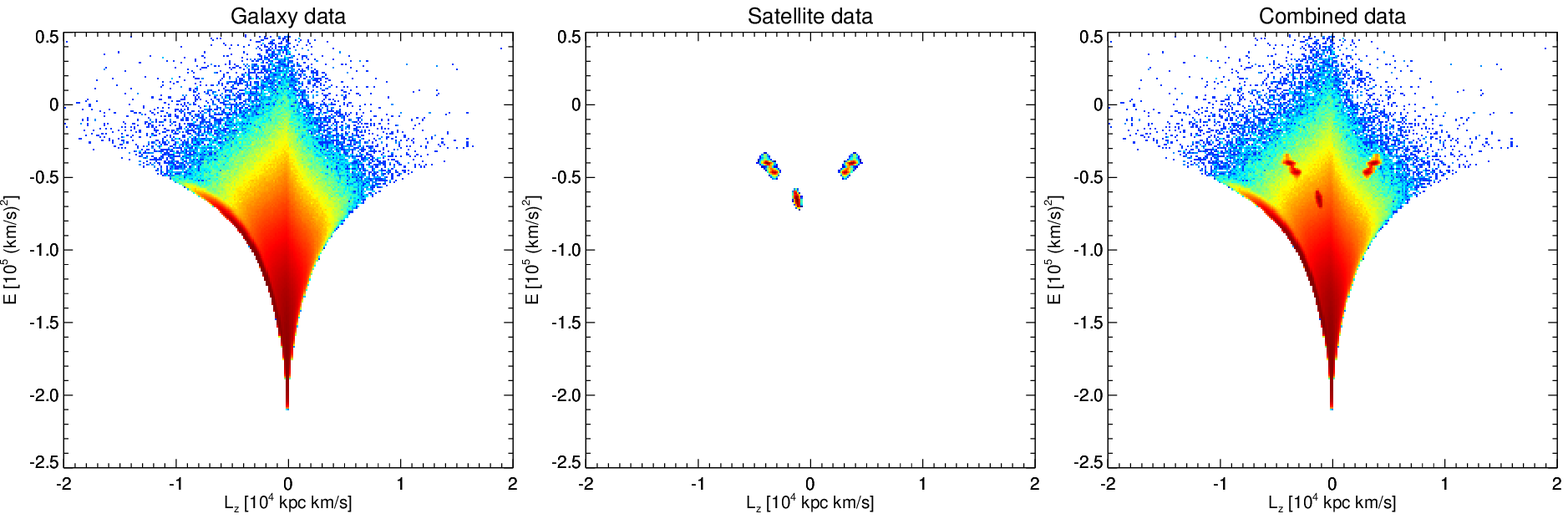}\\
\includegraphics[width=\textwidth]{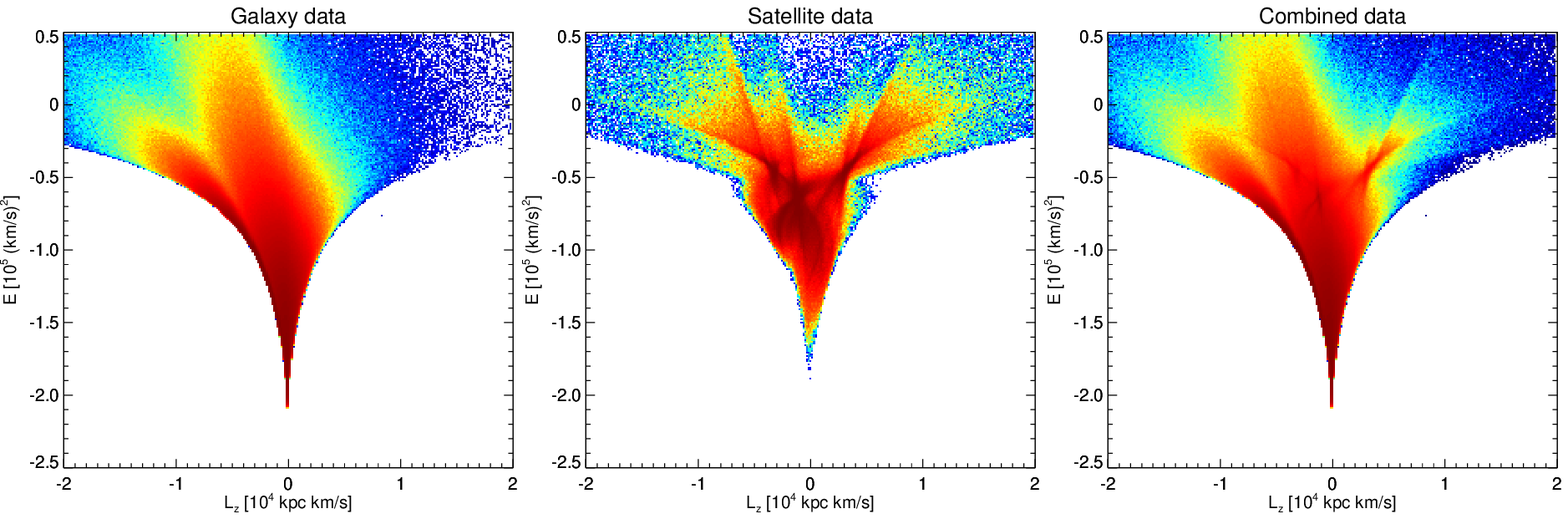}\\
\includegraphics[width=\textwidth]{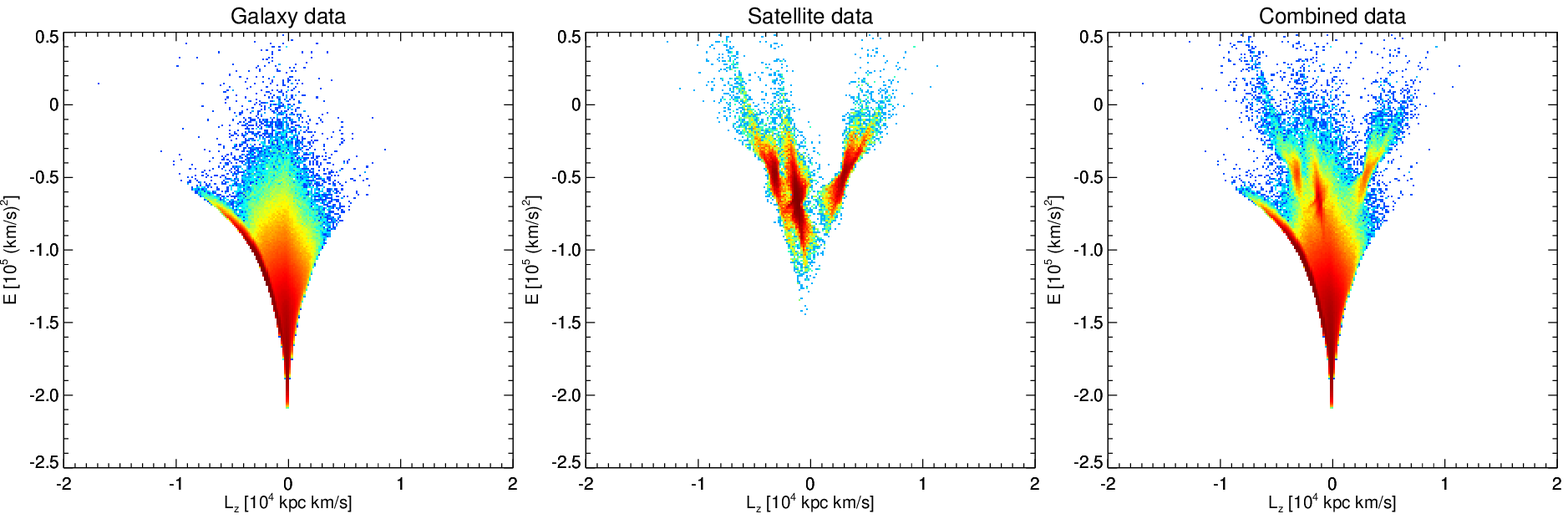}
\caption{$E$--$L_z$ space for three of our disrupted $N$-body satellites and
the synthetic Milky Way catalogue. The upper panels show the error free
$E$--$L_z$ diagrams (from left to right: Galaxy stars only, satellite stars
only, and Galaxy and satellite stars combined, respectively). The middle
panels show $E$--$L_z$ space after the expected \textit{Gaia} observational errors have
been added. Only stars with radial velocities and positive measured parallaxes
are plotted. Finally the lower panels show only the simulated stars from a
high-quality sample for which $V<15$ and $\varpi/\sigma_\varpi > 5$. These
figures have been constructed as two-dimensional histograms by counting stars
in 250$\times$250 cells that divide $E$--$L_z$ space. The white areas
correspond to empty cells while the individual dots indicate cells with 1 or
more stars. Areas where all contiguous cells are filled appear as a grey
scale.}
\label{fig:elzspace}
\end{figure*}

\subsection{Searching for satellite remnants in $E$--$L_z$ space}
\label{sec:elzsearch}

We constructed $E$--$L_z$ diagrams for all of our satellite $N$-body
simulations together with the synthetic catalogue of the background population
associated with the Milky Way. For clarity in the following discussion we
concentrate on just a few satellites. The diagrams are shown in
Fig.~\ref{fig:elzspace}. The upper panels refer to the $E$--$L_z$ diagrams
without \textit{Gaia} errors (from left to right: Galaxy only, satellites only and
Galaxy and satellites combined). The middle panels show the effect of adding
the expected \textit{Gaia} observational errors. In this case the sample was restricted
to stars with positive parallaxes and measured radial velocities in order to
calculate $E$ and $L_z$. This selection criterion introduces an implicit limit
on the apparent brightness of the stars of $V\la17$--$18$. In the bottom
panels the sample of stars selected from the catalogue is restricted to stars
brighter than $V=15$ having a parallax signal-to-noise of
$\varpi/\sigma_\varpi >5$ . In these diagrams, the entire Galaxy catalogue
consists of $3.5 \times 10^8$ stars and the satellite models correspond to
runs 1 and 5 of Table~\ref{tab:satorb} where a third satellite was generated
by reversing the angular momentum of the satellite from run 1. All satellites
have $M_{V,\mathrm{tracer}}\,=\,0.5$ (i.e. $M_{\mathrm{V,tot}} \approx -17$)
and the stellar luminosities and colours were computed using a $10$ Gyr
isochrone with metallicity $Z=0.004$ and a mass function $\xi(m) \propto
m^{-1.5}$.  Energies were computed employing the Galaxy model described in
section~\ref{sec:nbody}.

\begin{figure*}
\includegraphics[width=0.5\columnwidth]{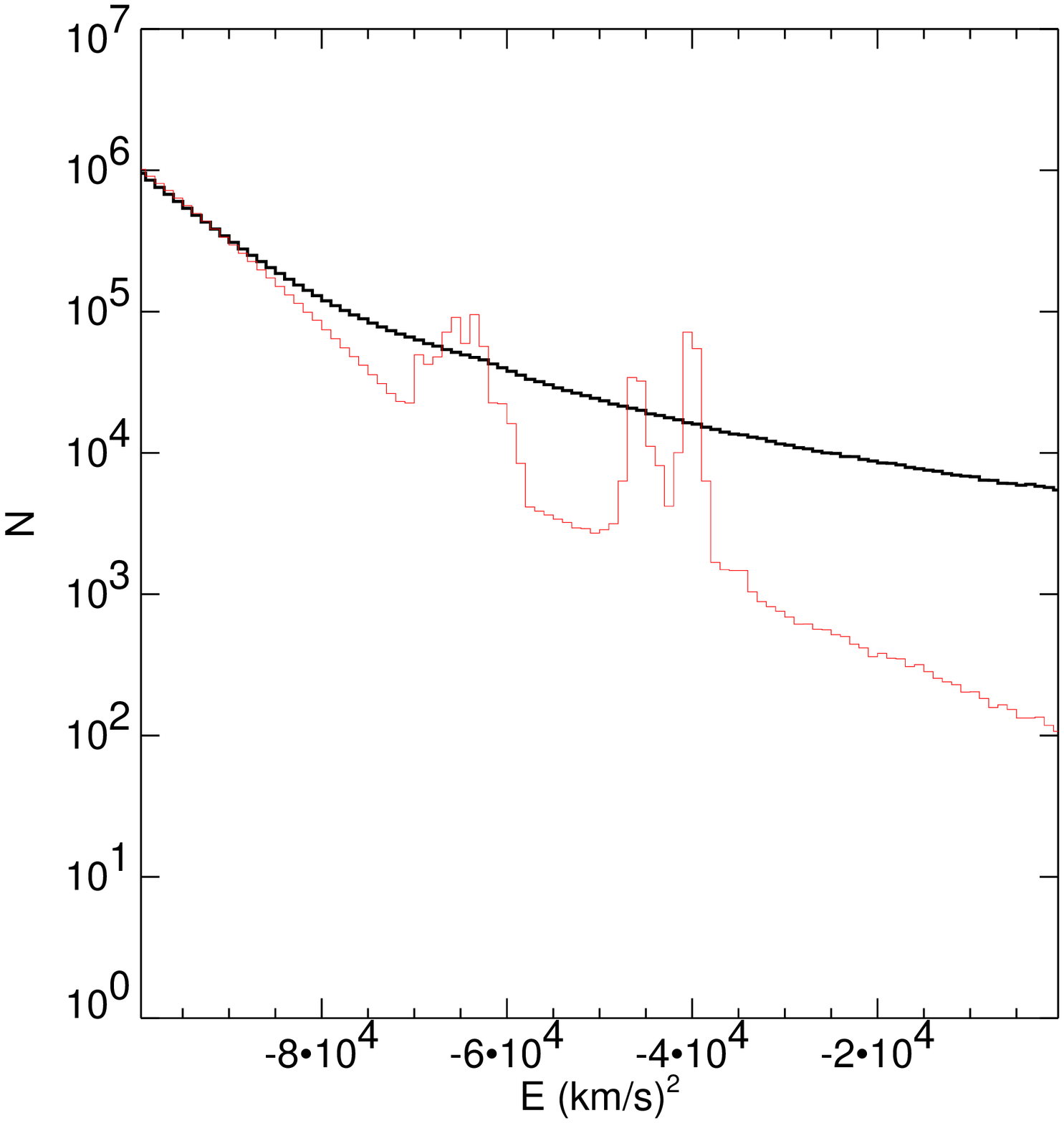}
\includegraphics[width=0.5\columnwidth]{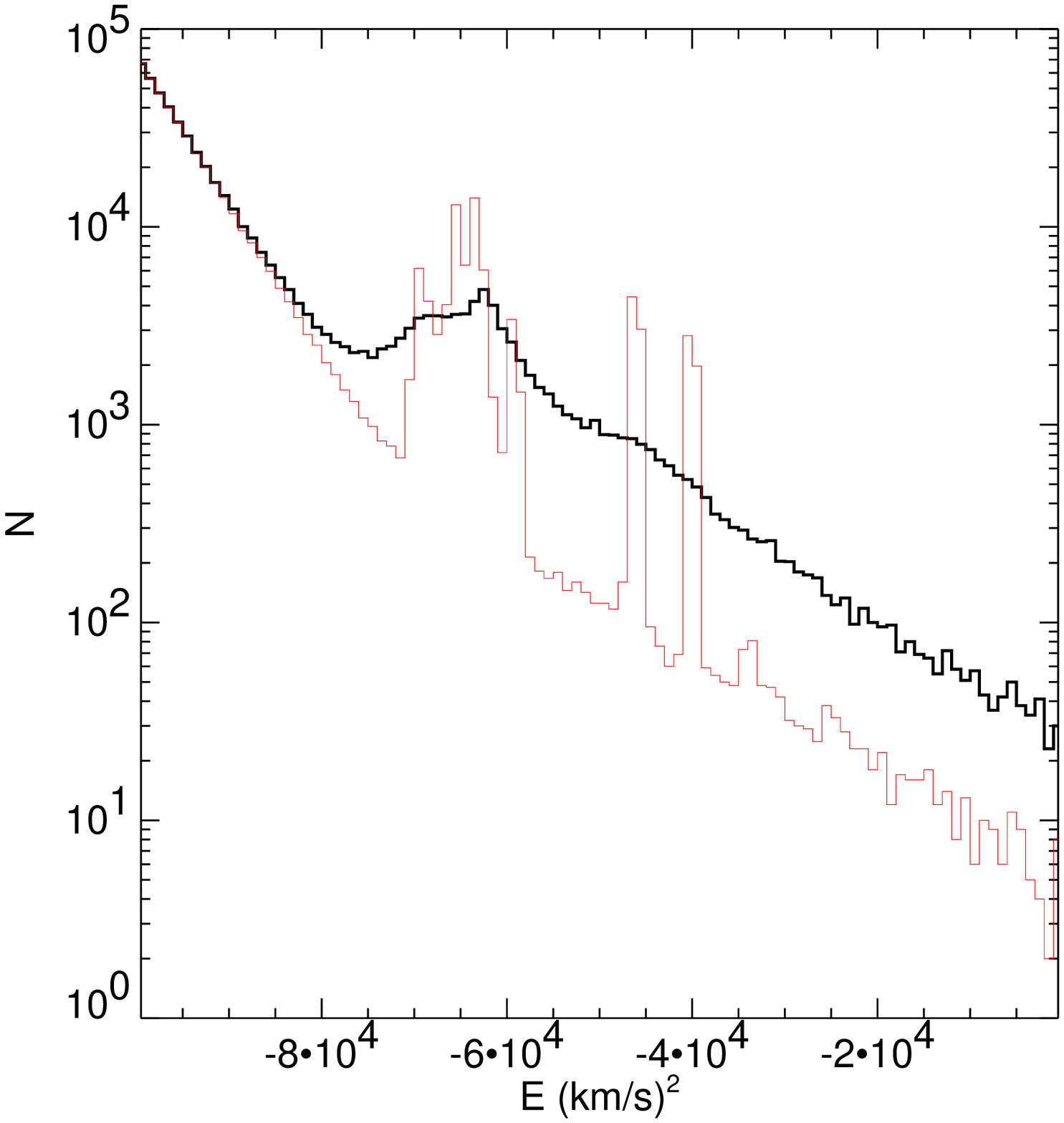}
\includegraphics[width=0.5\columnwidth]{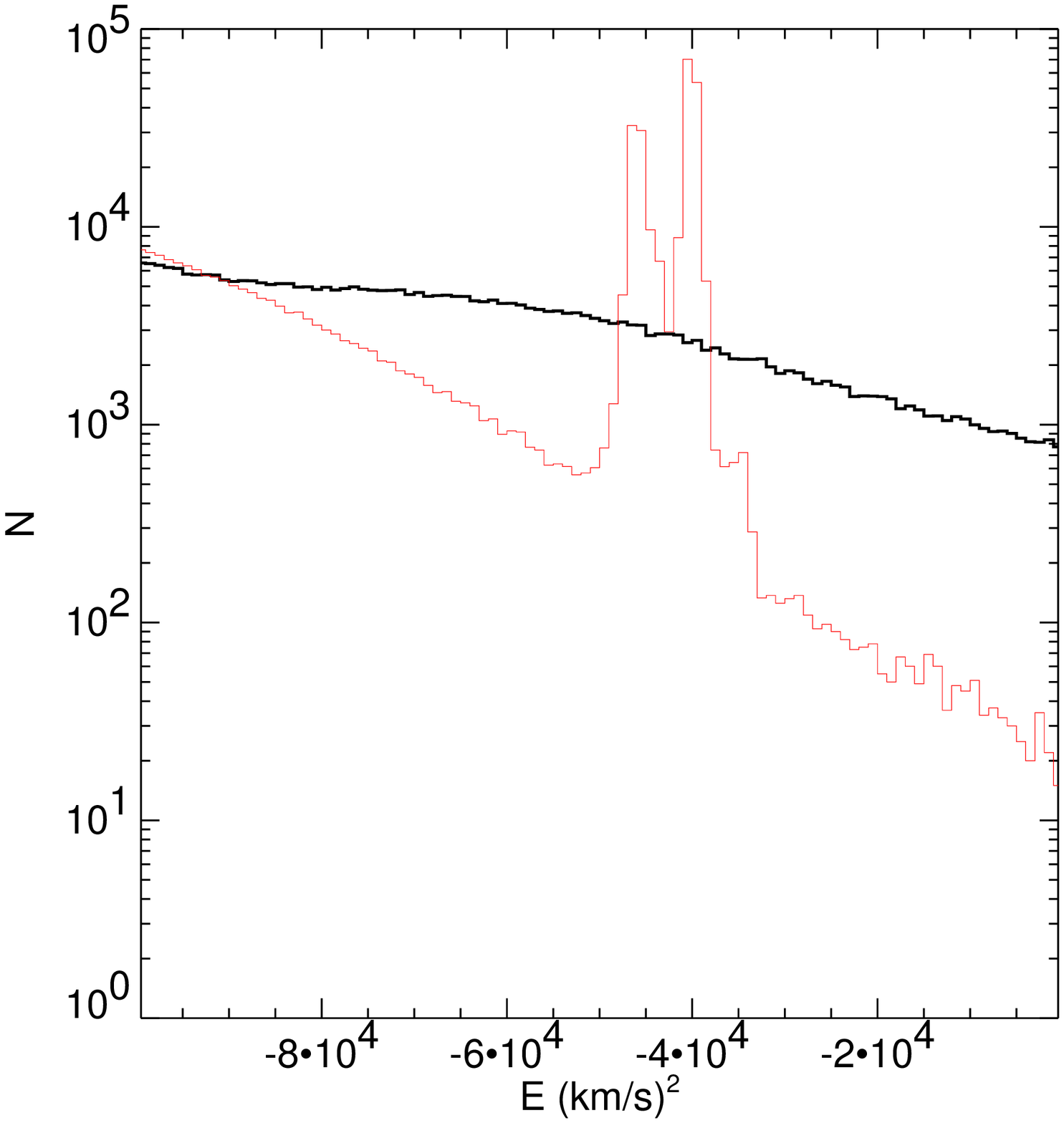}
\includegraphics[width=0.5\columnwidth]{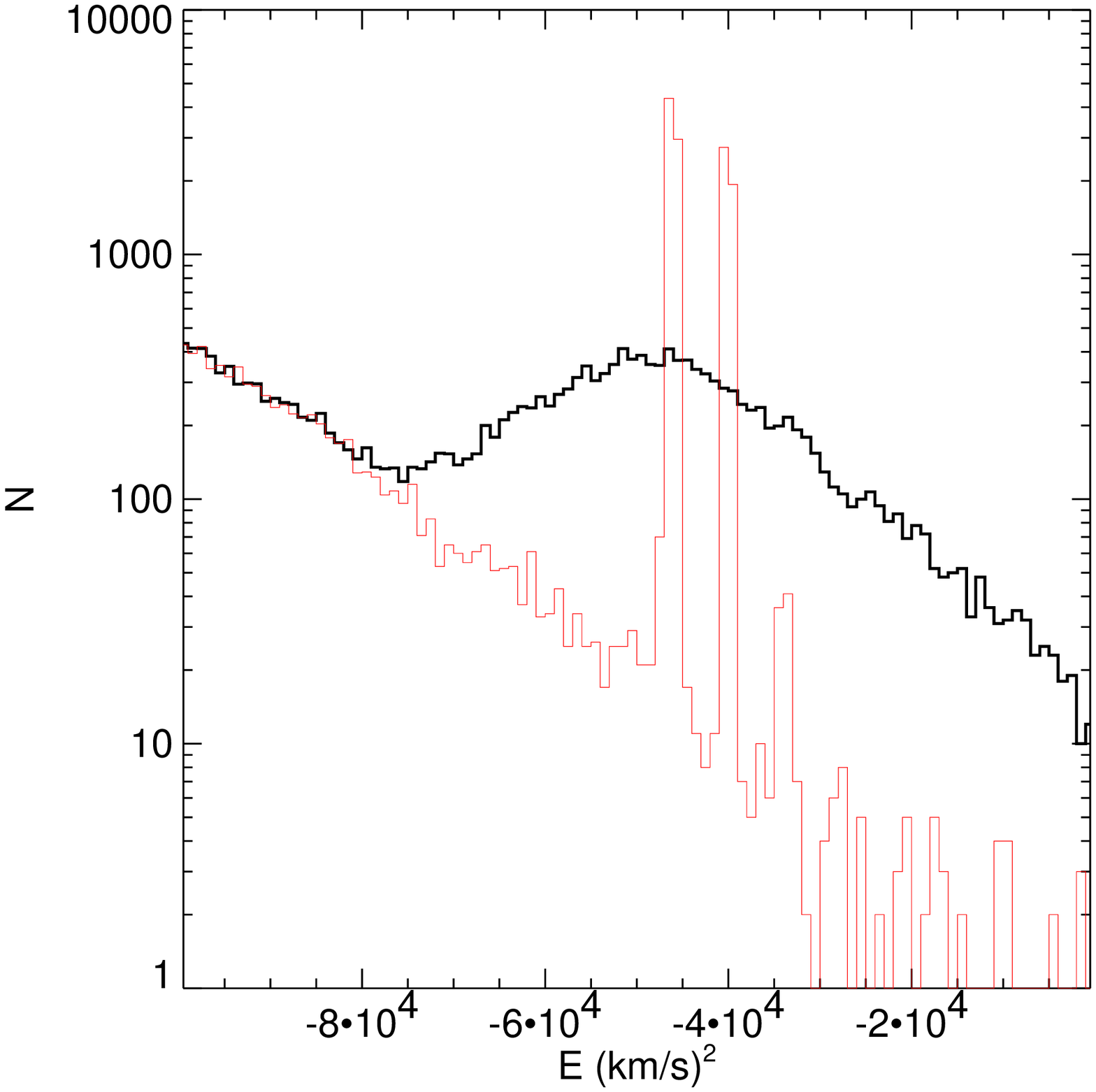}
\caption{ Energy histograms obtained from $E$--$L_z$ space for all satellite
and Galaxy stars by taking slices in the integral of motion $L_z$. The leftmost
two $E$-histograms correspond to all stars located in the interval
$-7\times10^3 < L_z< 0$ containing the prograde satellites (from left to right
the entire and restricted \textit{Gaia} samples, respectively). Similarly, the right
two $E$-histograms were constructed by using the interval $10^3< L_z < 7\times
10^3$ enclosing the retrograde satellite. In all cases, thin lines indicate
error-free data while thick lines correspond to data after \textit{Gaia} errors are
added. Prograde satellites are centred at energies $E \approx -6.5 \times
10^4$ (km~s$^{-1}$)$^2$, $E \approx -4.2\times 10^4$ (km~s$^{-1}$)$^2$; the retrograde
satellite is centred at energy $E \approx -4.2\times 10^4 $ (km~s$^{-1}$)$^2$.  }
\label{fig:Ehist}
\end{figure*}

In practice one will not know the actual potential of the Milky Way and thus
an estimate of this quantity is required to calculate $E$. To simulate the
effect of an erroneous estimate of the potential we also computed the energy
using an alternative Galaxy potential model \citep[see equations (8)--(10)
of][]{Helmi2000}. This results in a larger spread of the Galaxy and satellite
stars in energy and this spread can be used in combination with the angular
momentum measurements and a detailed modelling of the identified debris
streams in order to constrain the Galactic potential. However the $E$--$L_z$
diagrams with the observational errors included look largely the same as those
in Fig.~\ref{fig:elzspace} so we will not consider this issue further in the
discussion that follows.

It can be seen that satellite remnants are easy to identify if there are no
observational errors, even in the presence of a stellar background (upper
panels of Fig.~\ref{fig:elzspace}). However, once the observational errors are
introduced, the identification process becomes more difficult. Notice how the
satellite and Galaxy stars are now smeared out in preferential directions in
the $E$--$L_z$ diagram, which is caused by the dominance of the parallax
errors as illustrated in Fig.~\ref{fig:errorpropag}. The middle panels of
Fig.~\ref{fig:elzspace} demonstrate that if the entire \textit{Gaia} Survey is
taken into account it becomes very difficult to disentangle the different
merger events. In particular, satellites on prograde orbits become buried in
the very crowded region occupied by the Galactic disc. The identification
process can be significantly improved when the search in $E$--$L_z$ is
restricted to a `high quality' sample as illustrated in the lower panels of
Fig.~\ref{fig:elzspace}. In this case, even a visual identification clearly
shows all three satellites despite the presence of the Galactic stellar
background. In this respect, the middle diagram of the lower panels in
Fig.~\ref{fig:elzspace} is similar to Fig.~4 of \citet{Helmi2000}. However the
clumping of the debris streams in $E$--$L_z$ is less clear and they are
connected to each other by the background population of Milky Way stars. This
makes the application of a `friends of friends' algorithm to search for debris
streams \citep[as proposed by][]{Helmi2000} less attractive.

\begin{figure}
\includegraphics[width=0.5\columnwidth]{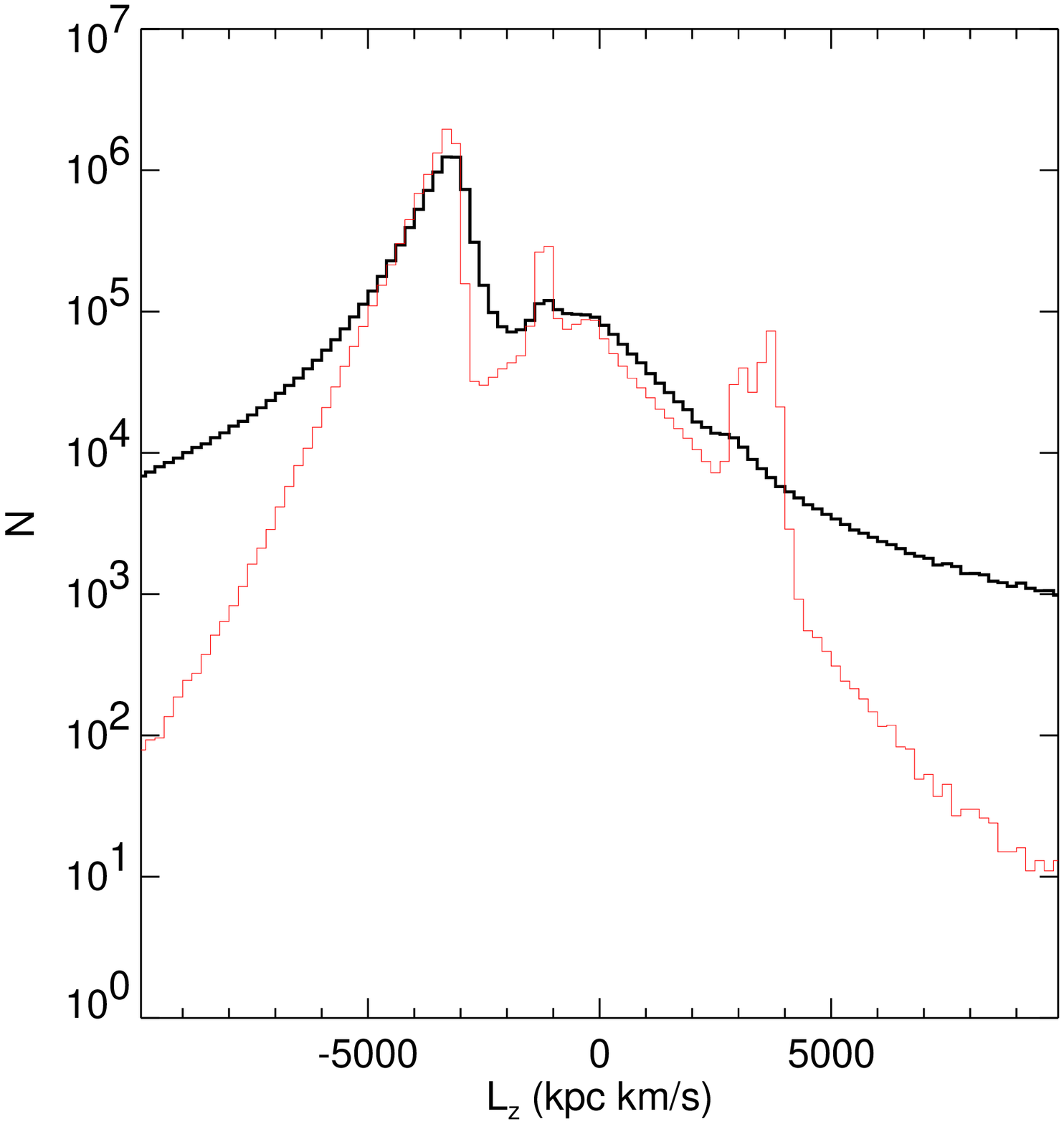}
\includegraphics[width=0.5\columnwidth]{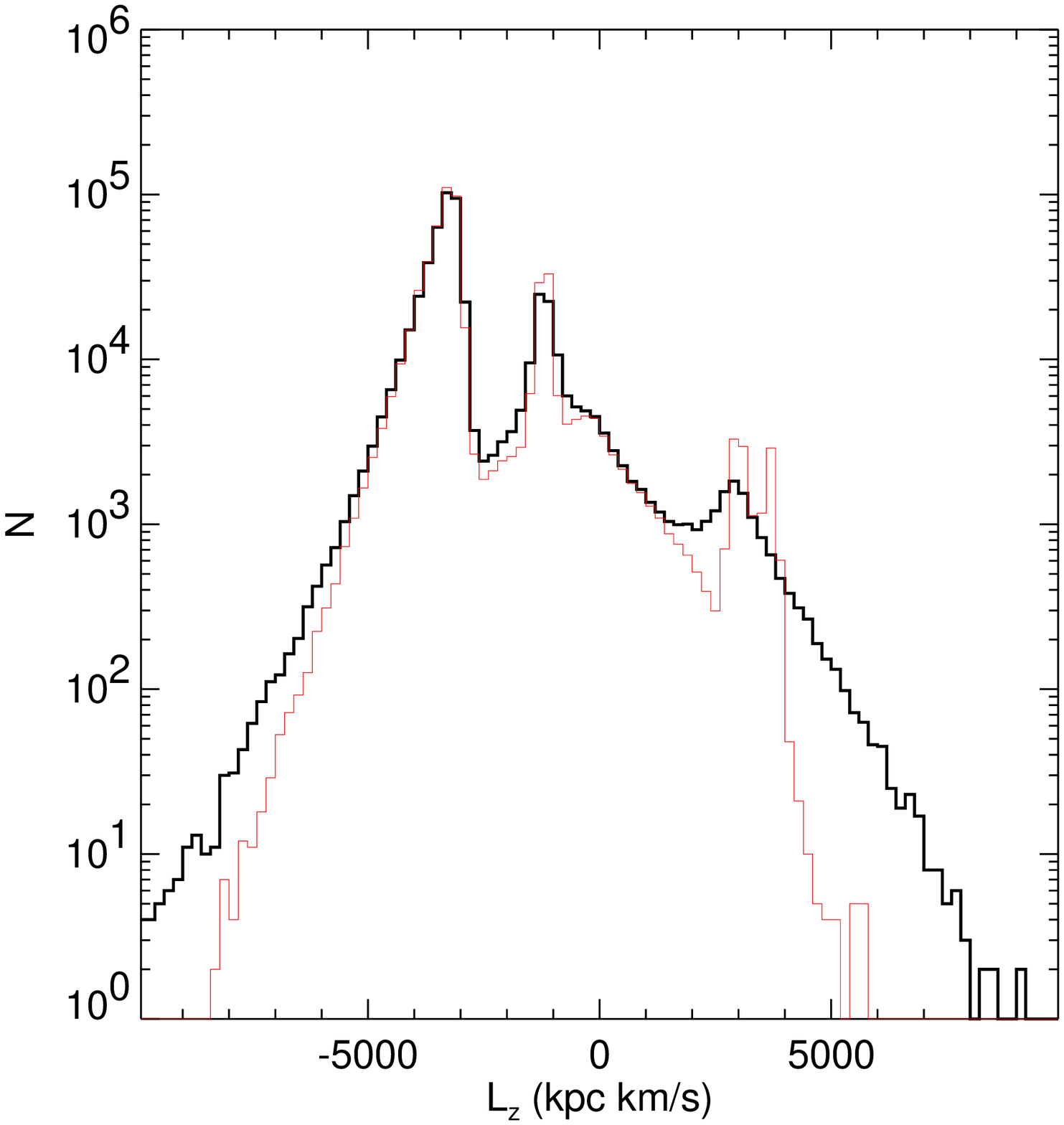}
\caption{$L_z$ histograms obtained from $E$--$L_z$ space for all satellite and
Galaxy stars by taking a slice in the integral of motion $E$. The left panel
shows the $L_z$-histogram considering the whole survey in the energy interval
$-10^5 < E < 0$ while the right panel shows the distribution of $L_z$ for our
high quality sample in the same energy range. The lines have the same meaning
as in Fig.~\ref{fig:Ehist}.}
\label{fig:Lzhist}
\end{figure}

Another very important feature of these $E$--$L_z$ diagrams is that the
observational errors introduce a lot of false high density structures
associated with the satellites (see middle column in
Fig.~\ref{fig:elzspace}). These structures look like singularities or caustics
in phase-space and from a naive analysis could be mistaken for features in the
topology of phase-space corresponding to physical entities or past merger
events that in reality do not exist.

Taking slices in the integrals of motion by selecting certain ranges of $E$ or
$L_z$ allows us to make a more quantitative assessment by focusing our
attention on $E$ or $L_z$-histograms containing the relics of the disrupted
satellites. A similar approach has been used by \citet{Meza2005} for the case
of the debris of the $\omega$Cen dwarf, the galaxy that presumably contributed
the $\omega$~Cen globular cluster to our galaxy.  They were able to identify
histogram peaks coinciding with the loss of satellite particles in the last
three pericentric passages (see their fig.~4). However, our case proves to be
more challenging, as can be appreciated in Figs.~\ref{fig:Ehist} and
\ref{fig:Lzhist}. Here the thin lines refer to error-free data and the thick
lines to data after \textit{Gaia} errors have been taken into account. $E$
histograms for error-free data show a double peak around each satellite. These
peaks seems to be related to stars coming from the leading and trailing tails
of the disrupted satellite.  However, these peaks are smeared out over the
entire energy space when \textit{Gaia} errors are added, while only some bumps
survive in our high quality sample (notice that even in this more favourable
situation any trace of one of the prograde satellites has been completely
erased). A similar behaviour is seen in the $L_z$ histograms, although in this
case all three satellites are preserved after the errors are added in the high
quality sample.

Finally, we note that dynamical friction has been ignored in our numerical
simulations of satellite disruption. If dynamical friction is taken into
account $L_z$ will not be conserved and its evolution will strongly depend on
the initial satellite mass as well as on the orbital angular momentum with
respect to the angular momentum of the main Galaxy. For example prograde
orbits decay faster than retrograde ones
\citep{Huang1997,Velazquez1999}. However, \citet*{HWS2003} have pointed out
that the conservation of phase-space density remains a promising technique for
providing useful information about the Galaxy formation history even in a more
realistic situation where an entire galaxy builds up from mergers in a
$\Lambda$-CDM cosmology.

\begin{figure*}
\includegraphics[width=0.33\textwidth]{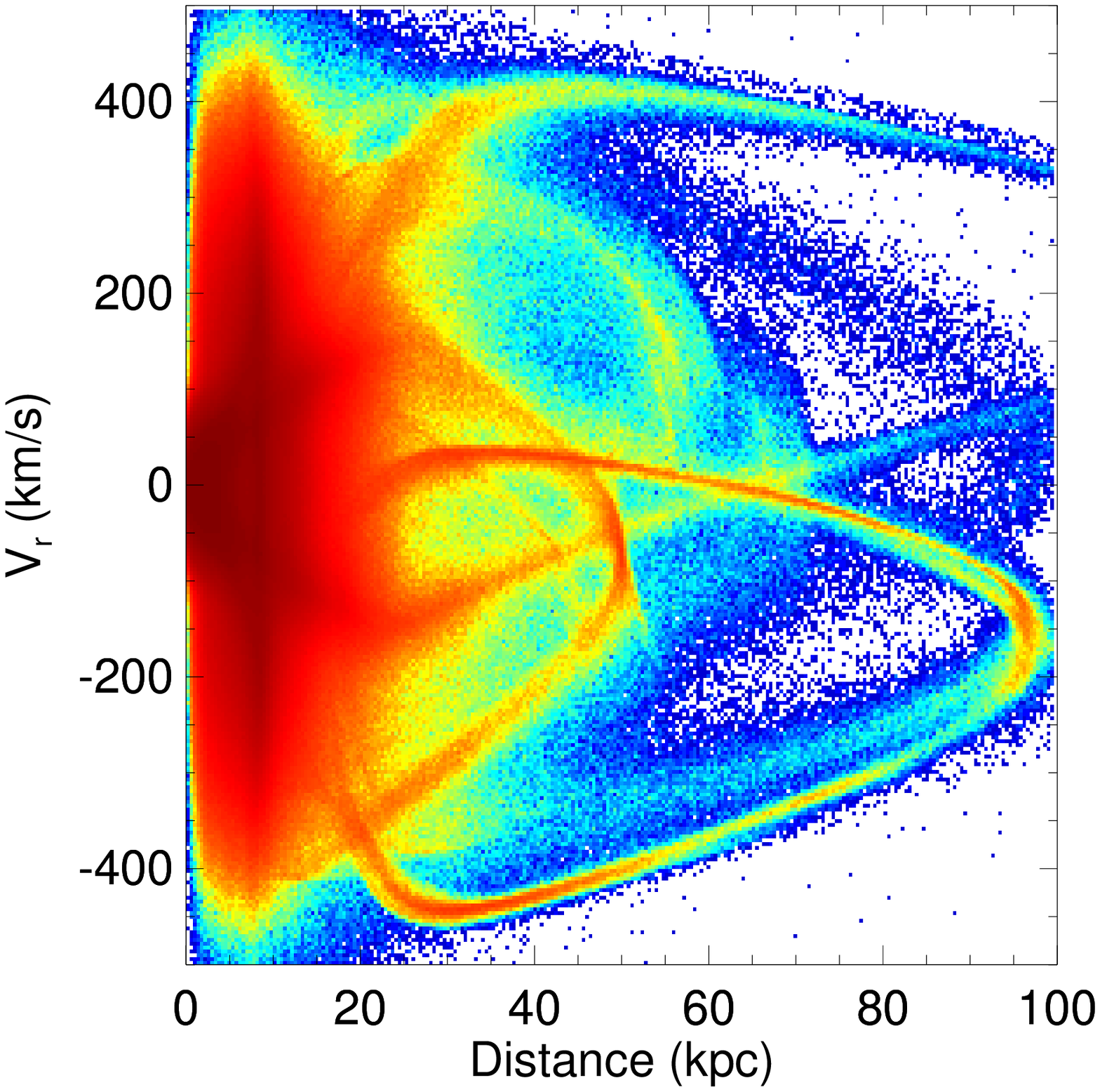}
\includegraphics[width=0.33\textwidth]{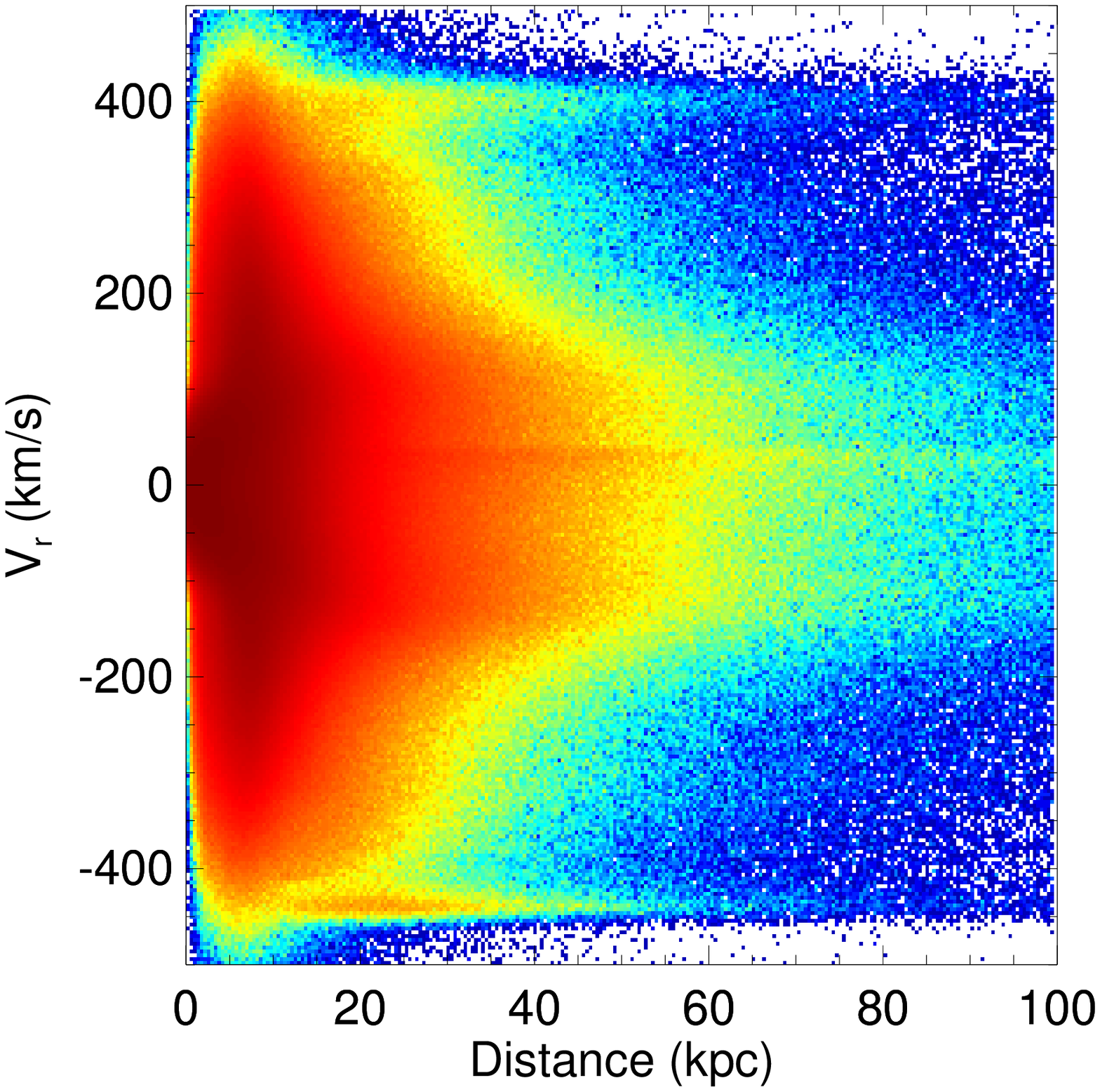}
\includegraphics[width=0.33\textwidth]{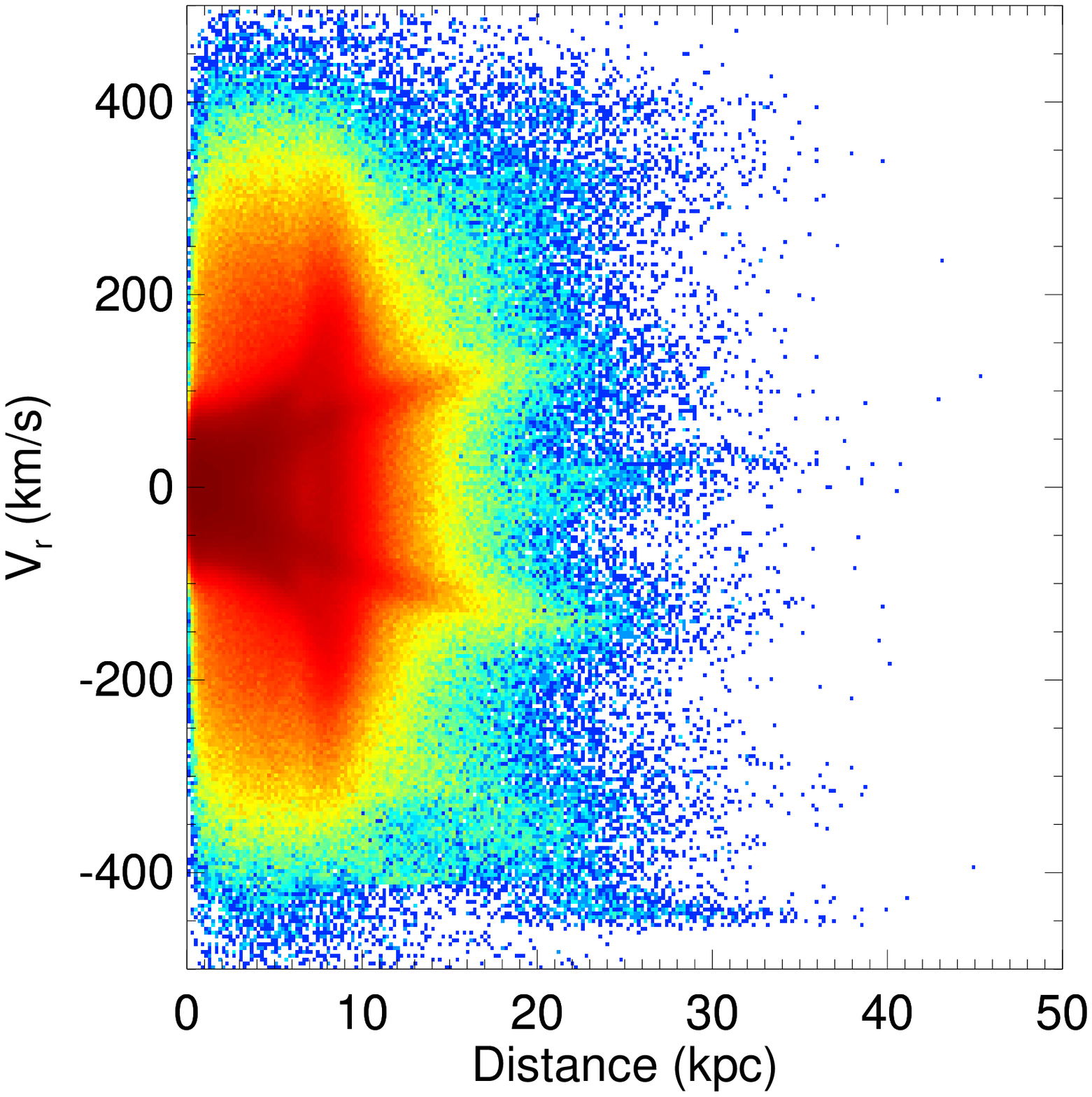}
\caption{Radial velocity versus distance diagrams for our simulated \textit{Gaia}
  catalogue. The same satellites as in Fig.~\ref{fig:elzspace} were used and
  superposed on the Galactic background. The left panels shows the data
  without observational errors. The middle panel shows the effect of adding
  the \textit{Gaia} measurement errors, and the right panel shows the \textit{Gaia} data for a
  high quality subsample restricted in magnitude, $V<15$, and parallax,
  signal-to-noise $\varpi/\sigma_\varpi > 5$ as in
  Fig.~\ref{fig:elzspace}. Note the change in the horizontal scale in this
  last panel. These figures were generated in the same way as
  Fig.~\ref{fig:elzspace}.}
\label{fig:rvr}
\end{figure*}

\subsection{Searching for satellite remnants in $r$--$v_\mathrm{rad}$ space}
\label{sec:rvrsearch}

\citet*{Johnston1995} suggested that diagrams of radial velocity versus
distance would provide a powerful tool to look for fossil signatures in the
halo of our Galaxy. These $r$--$v_\mathrm{rad}$ diagrams are related to the
$E$--$L_z$ diagrams since, roughly speaking, the total energy of particles
reflects the apocentre of the orbit while the pericentric radius is associated
with the orbital angular momentum. This technique has not been applied in
practice owing to a lack of data sets with enough radial velocities and
reliable distances.

Figure~\ref{fig:rvr} shows the $r$--$v_\mathrm{rad}$ diagrams for our simulated
\textit{Gaia} catalogue. The same satellites were used as for the $E$--$L_z$
diagrams. The data without the observational errors are shown in the left
panel. The debris of the disrupted satellites can clearly be distinguished as
narrow structures in this diagram but it is not possible to associate the
latter with one or another of the satellites. When the observational errors
are included most of these fossil signatures are washed out and a
straightforward detection will be quite difficult as can be appreciated in the
middle $r$--$v_\mathrm{rad}$ diagram. Restricting the data to the same high
quality subsample as defined above, does not change the situation although it
is possible to discern some structures related to the tidal debris left by the
disrupted satellites.

\subsection{The need for complementary search methods}

These results show that a straightforward search for the debris of satellite
galaxies in $E$--$L_z$ or $r$--$v_\mathrm{rad}$ space may not be very
efficient in practice. A good start can be made by limiting the search to high
quality samples, where the contrast between satellites and Galaxy can be
further enhanced by concentrating on certain regions of the sky (away from the
Galactic disc for example). However the effects of error propagation and
selection biases (see Section~\ref{sec:elzerrors}) should be carefully
accounted for. Once a first set of stars belonging to a particular remnant has
been identified in $E$--$L_z$ one can in principle trace the rest of the
stream by modelling the orbit of the satellite and by making use of the
astrophysical properties of the stars (age, metallicity, alpha-element
enhancements) which will be known from the photometric data provided by
\textit{Gaia}. Alternatively one can select samples of metal poor stars
(thereby also lowering the Galaxy-satellite contrast) from the \textit{Gaia}
catalogue in order to study the very oldest remnants. \textit{Gaia} will also
provide extremely well calibrated photometric distances based on the
parallaxes of the nearer and/or brighter stars. These calibrations can be used
to obtain accurate distances for faint stars for which \textit{Gaia}
parallaxes are of poor quality. Better distances for the fainter stars will
dramatically improve the quality of the $E$--$L_z$ and $r$--$v_\mathrm{rad}$
diagrams. These complementary procedures can be studied with our simulations
by adding a more complete model of the \textit{Gaia} photometric data.

%%%%%%%%%%%%%%%%%%%%%%%%%%%%%%%%%%%%%%%%%%%%%%%%%%%%%%%%%%%%%%%%%%%%%%%%%%%%%%%

\section{Conclusions}
\label{sec:future}

\textit{Gaia} will provide us with an unprecedented wealth of information on
the structure and formation history of the Milky Way galaxy, for the first
time providing full phase-space information across its entire
volume. Exploiting this data set for studies of the assembly of the halo and
other components of the Milky Way will be challenging owing to the enormous
volume of data; astrometry and photometry for 1 billion stars and radial
velocities for 100--150 million stars, over the entire sky. A good preparation
will require familiarizing ourselves with the handling of such large datasets
as well as exploring the best ways of disentangling remnants of past merger
events from each other and from the Galactic background population.

We have started to address this problem by simulating the \textit{Gaia} catalogue with
a realistic number ($10^8$--$10^9$) of entries. This was done by combining a
Monte Carlo model of the Milky Way with $N$-body simulations of satellites
being disrupted as they orbit in the potential of our Galaxy. The Monte Carlo
model of the Milky Way consists of $3.5\times10^8$ stars and was generated by
taking into account how the \textit{Gaia} survey limit introduces magnitude limited
spheres for each stellar type. The region on the sky limited by Galactic
coordinates $-90^\circ\leq\ell\leq +90^\circ$ and $-5^\circ\leq b\leq
+5^\circ$ was left out of the simulations for practical (computation time)
reasons.

The models of the debris streams were generated by simulating the disruption
of dwarf galaxies orbiting our galaxy. The Milky Way is represented by a rigid
potential and a tree-code was used to carry out the $N$-body simulations of
the orbiting satellites over a time period of about 10~Gyr. The simulations
were carried out for satellites placed on five different orbits which vary in
apocentre, pericentre and the initial inclination with respect to the Milky
Way's disc. Multiple debris streams can then be simulated by combining
different $N$-body snapshots (different orbits and/or ages), and each snapshot
can be rotated around the $Z$-axis, flipped with respect to the disc plane, or
reversed in angular momentum.

We describe in detail how to correctly combine the Milky Way and satellite
models such that the variation of the number of visible stars along the orbit
of a satellite is taken into account. This is most easily achieved by assuming
that the $N$-body particles represent a bright tracer population of a larger
underlying dwarf galaxy.

After combining the models the synthetic \textit{Gaia} catalogue was generated by
adding the astrometric and radial velocity errors according to the
prescriptions resulting from detailed preparatory studies of the \textit{Gaia}
mission. The observational errors were modelled taking into account the
dependence on apparent magnitude, colour and sky position of the stars. To
assign photometric properties to the particles we use a Hess diagram for the
Solar neighbourhood for Galactic particles, while for the satellite particles
we use isochrones from the Padova group. Our work thus introduces a thorough
methodology to incorporate $N$-body simulations within a simulation of an
actual observational survey in a manner that addresses questions of
kinematics, metallicity and photometric properties of the simulated system, as
well as errors in the observables.

The catalogue was used to study the appearance of the Galaxy and satellite
data in $E$--$L_z$ space. The results show that it will be challenging to trace
the debris streams throughout the entire volume of the Galaxy when \textit{Gaia} errors
are properly accounted for. However, limiting the sample to bright stars with
good parallaxes allows a preliminary search to be done which should be
followed up by complementing searches in $E$--$L_z$ by, for example, a further
tracing of the debris streams based on the astrophysical properties of the
stars which will be known from the photometry provided by \textit{Gaia}. In addition,
using photometric distances, which will be very well calibrated after the \textit{Gaia}
mission, will yield accurate distances for faint stars which can be used
instead of the parallaxes, thus avoiding the problems associated with the
propagation of parallax errors into the integrals of motion.

Ultimately our aim is to use these simulations to study much more thoroughly
and realistically the possible search techniques one can use to recover the
remnants of satellites in the stereoscopic and multi-colour data set that \textit{Gaia}
will provide. This includes a more complete study of the use of the integrals
of motion method by also using the full angular momentum vector $L$ and a
further exploration of the radial velocity versus distance diagrams. In addition
we will explore how the astrophysical information from the \textit{Gaia} photometry can
be used in the search for debris streams. This requires a number of
improvements to our simulations, which include:
\begin{itemize}
\item A proper model of the \textit{Gaia} photometric system which is available using
  the tools developed by the \textit{Gaia} Photometric Working Group
  \citep[see][]{Jordi2005,Carrasco2005}.
\item Updating the predicted \textit{Gaia} observational errors to the latest available
  assessments. Specifically, the radial velocity errors we used are somewhat
  optimistic and a better assessment of them is now available
  \citep{Katz2004}.
\item Using a clumped distribution of stars in the halo. The simulation of a
  clumpy halo will not be trivial and will probably require simulating
  multiple merger events which combine to form the halo. This may be obtained
  from the results of detailed cosmological simulations.
\item Including an extinction model and possibly adding a thick disc to the
  Galaxy model. Within the \textit{Gaia} project 3D extinction models have been
  developed which can be added to our simulations \citep[see][]{Drimmel2005}.
\item Using more sophisticated models for the dwarf galaxies. The $N$-body
  simulations should include dynamical friction and possibly more
  particles. The simulation of the stellar properties of the satellite
  particles should be updated with a more realistic mass function and possibly
  with the inclusion of more than one stellar population in the dwarf
  galaxy. The need for the latter is supported by the recent results
  concerning the stellar population of the Sculptor dwarf galaxy in which two
  distinct ancient stellar components were found \citep{Tolstoy2004}.
\end{itemize}

%%%%%%%%%%%%%%%%%%%%%%%%%%%%%%%%%%%%%%%%%%%%%%%%%%%%%%%%%%%%%%%%%%%%%%%%%%%%%%%%

\section*{Acknowledgements}

AGAB thanks everyone at the Instituto de Astronom{\'\i}a in Ensenada for their
hospitality during two visits in which most of the work described here was
done. HMV and LAA acknowledge support from DGAPA/UNAM grants IN113403 and
IN111803 as well as CONACyT grant 27678--E. Figure~\ref{fig:skyproject} was
made using the HEALPIX\footnote{http://www.eso.org/science/healpix}
\citep{Gorski1999} package.

%%%%%%%%%%%%%%%%%%%%%%%%%%%%%%%%%%%%%%%%%%%%%%%%%%%%%%%%%%%%%%%%%%%%%%%%%%%%%%%%

%%%%%%%%%%%%%%%%%%%%%%%%%%%%%%%%%%%%%%%%%%%%%%%%%%%%%%%%%%%%%%%%%%%%%%%%%%%%%%%%

\appendix
\section{Transforming Phase-Space Coordinates to \textit{Gaia} Data}
\label{ap:gaiamod}

Here we describe in detail how the phase-space coordinates
$(\vect{r},\vect{v})$ of our model Galaxy were converted to the data in the
simulated \textit{Gaia} catalogue; the five astrometric parameters $\ell$, $b$, $\varpi$,
$\mu_{\ell*}$, and $\mu_b$ and the radial velocity $v_\mathrm{rad}$.  The
barycentric position $\vect{r}^\mathrm{b}=\vect{r}-\vect{r}_\odot$ and
velocity $\vect{v}^\mathrm{b}=\vect{v}-\vect{v}_\odot$ are related to the
astrometric and radial velocity data as follows \citep[see e.g.][Sections 1.2
and 1.5]{ESA1997}:
\begin{equation}
\vect{r}^\mathrm{b} = A_p\uvect{r}/\varpi
\end{equation}
and:
\begin{equation}
\vect{v}^\mathrm{b} = \uvect{p}\mu_{\ell*}A_v/\varpi + \uvect{q}\mu_b
A_v/\varpi + \uvect{r}v_\mathrm{rad}\,,
\end{equation}
where $[\uvect{p}\ \uvect{q}\ \uvect{r}]$ is the normal triad defined below,
and $A_p=1000$~mas~pc and $A_v=4.74047...$~yr~km~s$^{-1}$ designate the
astronomical unit in the appropriate form. Note that if the distances are in
units of kpc, the units of parallax and proper motion are micro-arcsecond
($\mu$as) and micro-arcsecond per year ($\mu$as~yr$^{-1}$).

In the galactic coordinate system the components of the normal triad
$[\uvect{p}\ \uvect{q}\ \uvect{r}]$ are given by:
\begin{eqnarray}
\mat{R} &=&  \left( \begin{array}{ccc}
               p_x&q_x&r_x\\ p_y&q_y&r_y\\ p_z&q_z&r_z
               \end{array}\right) \nonumber \\
                   &=&
\left(
\begin{array}{ccc}
-\sin\ell & -\sin b\cos\ell & \cos b\cos\ell \\
\phantom{-}\cos\ell & -\sin b\sin\ell & \cos b\sin\ell \\
 0 & \cos b & \sin b
\end{array}
\right) \,,
\end{eqnarray}
where \uvect{r} is the unit vector that specifies the (instantaneous)
barycentric coordinate direction $(\ell,b)$\footnote{Note that the vector
\uvect{r} and the barycentric coordinate direction are not strictly the same,
the latter includes the effect of stellar proper motion. However as we are
here only interested in the instantaneous positions and motions of the stars
at the central catalogue epoch we need not concern ourselves with this
distinction\citep[for more details see][Vol.\ 1 Sections 1.2 and
1.5]{ESA1997}}, and \uvect{p} and \uvect{q} are the unit vectors in the
directions of increasing $\ell$ and $b$ at \uvect{r}. The galactic coordinate
components of $\vect{r}^\mathrm{b}$ and $\vect{v}^\mathrm{b}$ may thus be
written:
\begin{equation}
\left(
\begin{array}{c}
r^\mathrm{b}_x \\ r^\mathrm{b}_y \\ r^\mathrm{b}_z
\end{array} \right) = \mat{R}
\left(
\begin{array}{c}
0 \\ 0 \\ A_p/\varpi
\end{array} \right)
\label{eq:partopos}
\end{equation}
and:
\begin{equation}
\left(
\begin{array}{c}
v^\mathrm{b}_x \\ v^\mathrm{b}_y \\ v^\mathrm{b}_z
\end{array} \right) = \mat{R}
\left(
\begin{array}{c}
\mu_{\ell*}A_v/\varpi \\ \mu_b A_v/\varpi \\ v_\mathrm{rad}
\end{array} \right)
\label{eq:muvradtovel}
\end{equation}

These equations can be inverted to calculate the astrometric and radial
velocity data from the barycentric positions and velocities, yielding:
\begin{equation}
\varpi=A_p/|\vect{r}^\mathrm{b}|
\end{equation}
and
\begin{equation}
\left(
\begin{array}{l}
v_{\ell*} \\ v_b \\ v_\mathrm{rad}
\end{array} \right) = \mat{R}^\prime
\left(
\begin{array}{c}
v^\mathrm{b}_x \\ v^\mathrm{b}_y \\ v^\mathrm{b}_z
\end{array} \right) \,,
\end{equation}
where $v_{\ell*}=\mu_{\ell*}A_v/\varpi$ and $v_b=\mu_bAv/\varpi$ are the
velocity components perpendicular to the line of sight to the star, and
$\mat{R}^\prime$ indicates the transpose of \mat{R}.

\begin{table}
\caption{Radial velocity precision (in km~s$^{-1}$) as a function of stellar
  spectral type and magnitude $V$. These are the numbers used in our
  simulations, taken from Fig.~8.14 in \protect\citet{ESA2000}.}
\label{tab:vraderrs}
\begin{tabular}{crcr}
\hline
\multicolumn{2}{c}{OBA} & \multicolumn{2}{c}{FGKM} \\[2pt]
$V$ & $\sigma_{v_\mathrm{rad}}$ & $V$ & $\sigma_{v_\mathrm{rad}}$ \\
\hline
10 & 0.25 & 10 & 0.1 \\
15 & 4    & 16 & 1   \\
16 & 10   & 17 & 2   \\
17 & 50   & 18 & 6   \\
\hline
\end{tabular}
\end{table}

Once the astrometric data have been calculated from \vect{r}$^\mathrm{b}$ and
\vect{v}$^\mathrm{b}$ they are transformed to the Ecliptic coordinate system
before adding the astrometric errors. This is done to simulate the fact that
the astrometric errors will vary over the sky as a function of the Ecliptic
latitude $\beta$. As a consequence, upon rotation back to the Galactic
reference frame, correlations are introduced between the errors in $\ell$ and
$b$, and $\mu_{\ell*}$ and $\mu_b$. The transformation to the Ecliptic
coordinate system is done through a rotation matrix as explained in Section
1.5.3 of the introductory volume to the Hipparcos and Tycho Catalogues
\citep{ESA1997}.

The errors in the astrometry are then added as follows
\citep[see][]{Perryman2002,ESA2000}. The error on the parallax is given by:
\begin{eqnarray}
\sigma_\varpi \simeq &
(7+105z+1.3z^2+6\times10^{-10}z^6)^{1/2} \nonumber \\
 & \times[0.96+0.04(V-I)] \,,
\end{eqnarray}
where $z=10^{0.4(G-15)}$ and $G$ is the broad band \textit{Gaia} magnitude (defined by
the mirror reflectivity and CCD quantum efficiency curves) which can be
calculated from the approximate transformation:
\begin{eqnarray}
G = & V+0.51-0.50\times\sqrt{0.6+(V-I-0.6)^2} \nonumber \\
    & -0.065\times(V-I-0.6)^2 \,.
\end{eqnarray}
For the mean position and proper motion errors $\sigma_0$ and $\sigma_\mu$ the
following mean relations are used:
\begin{equation}
\sigma_0=0.87\sigma_\varpi \,,\quad \sigma_\mu=0.75\sigma_\varpi
\end{equation}
Finally, the variations of the errors as a function of $\beta$ are taken from
Table 8.3 in \citet{ESA2000}.

The errors on the radial velocity have been taken from fig.~8.14 in
\citet{ESA2000}, which shows the radial velocity accuracy for a number of
values of $V$ for both hot (OBA) and cool (FGKM) stars. The numbers are given
in Table~\ref{tab:vraderrs} and interpolation was used for intermediate values
of $V$. For OBA stars fainter than $V=17$ and FGKM fainter than $V=18$ no
radial velocities are generated.

Finally, the predicted observational errors for \textit{Gaia} as quoted here
reflect the estimates at the time the \textit{Gaia} Concept and Technology
Study Report was published \citep{ESA2000}. The values will evolve along with
changes in the details of the mission design and implementation.

\section{Angular momentum from astrometric and radial velocity data}
\label{ap:calclz}

Because the Sun is not located at the centre of the Galaxy the relation
between the angular momentum of a star and its measured position, parallax,
proper motion and radial velocity is complicated by the necessity to transform
from the barycentric to the Galactocentric frame. The angular momentum vector
$\vect{L}$ is calculated as
$\vect{r}\times\vect{v}=(\vect{r}^\mathrm{b}+\vect{r}_\odot)\times(\vect{v}^\mathrm{b}+\vect{v}_\odot)$,
which when worked out using Eqs.(\ref{eq:partopos}) and (\ref{eq:muvradtovel})
yields:
\begin{equation}
\vect{L} = \vect{L}^\mathrm{g} + \vect{L}^\odot\,,
\end{equation}
where $\vect{L}^\mathrm{g}$ is given by:
\begin{equation}
  \vect{L}^\mathrm{g} = \frac{A_pA_v}{\varpi^2}
  \begin{pmatrix}
    \phantom{-}\mu_b\sin\ell-\mu_{\ell*}\sin b\cos\ell \\[5pt]
    -\mu_b\cos\ell-\mu_{\ell*}\sin b\sin\ell \\[5pt]
    \mu_{\ell*}\cos b
  \end{pmatrix} \,,
\end{equation}
and $\vect{L}^\odot$ is given by:
\begin{equation}
  \vect{L}^\odot =
  \begin{pmatrix}
    -V_\odot\frac{A_p\sin b}{\varpi} \\[5pt]
    -R_\odot\left[\frac{A_v\mu_b\cos b}{\varpi}+v_\mathrm{rad}\sin b\right] \\[5pt]    
    R_\odot\Lambda+\frac{A_p\cos b\cos\ell}{\varpi}V_\odot
  \end{pmatrix} \,,\\
\end{equation}
where the term $\Lambda$ is given by:
\begin{equation}
\begin{split}
  \Lambda = & \frac{A_v\mu_{\ell*}\cos\ell}{\varpi}-
    \frac{A_v\mu_b\sin b\sin\ell}{\varpi}\\
    & +v_\mathrm{rad}\cos b\sin\ell+V_\odot \,.
\end{split}
\end{equation}
$\vect{L}^\mathrm{g}$ is the `Galactocentric' angular momentum which has no
physical meaning but corresponds to the angular momentum calculated by an
observer at the Galactic centre. The vector $\vect{L}^\odot$ indicates the
components of the observed angular momentum that arise from the transformation
from a barycentric to Galactocentric frame of reference. The norm of
$\vect{L}^\mathrm{g}$ correctly evaluates to:
\begin{equation}
|\vect{L}^\mathrm{g}|=\frac{A_pA_v}{\varpi^2}\sqrt{\mu_{\ell*}^2+\mu_b^2}\,,
\end{equation}
i.e., the distance to the Galactic centre multiplied by the total tangential
velocity. The equation above shows how the radial velocity enters the angular
momentum measurements and can be used to further study the way the astrometric
errors propagate in the measured angular momentum.

The equations for the energy of a star can be worked out similarly but they
depend on the model for the potential energy and hence we do not provide them
here.
\end{document}